\documentclass[graybox]{svmult}

\usepackage{mathptmx}       % selects Times Roman as basic font
\usepackage{helvet}         % selects Helvetica as sans-serif font
\usepackage{courier}        % selects Courier as typewriter font
\usepackage{makeidx}         % allows index generation
\usepackage{graphicx}        % standard LaTeX graphics tool
\usepackage{multicol}        % used for the two-column index
\usepackage[bottom]{footmisc}% places footnotes at page bottom

\makeindex             % used for the subject index
\begin{document}

\title{Fano resonances in plasmonic core-shell particles and the Purcell effect}
\author{Tiago Jos\'e Arruda, Alexandre Souto Martinez, Felipe A. Pinheiro, Romain Bachelard, Sebastian Slama, and Philippe Wilhelm Courteille}
\authorrunning{Tiago Jos\'e Arruda et al.}
\institute{
T. J. Arruda 
\at Instituto de F\'isica de S\~ao Carlos (IFSC), Universidade de S\~ao Paulo (USP), 13566-590 S\~ao Carlos, S\~ao Paulo, Brazil, 
\email{tiagojarruda@gmail.com}
\and
A. S. Martinez \at
  Faculdade de Filosofia,~Ci\^encias e Letras de Ribeir\~ao Preto (FFCLRP), Universidade de S\~ao Paulo (USP), 14040-901 Ribeir\~ao Preto, S\~ao Paulo, Brazil,
\email{asmartinez@usp.br }
\and
F. A. Pinheiro \at
	Instituto de F\'{i}sica, Universidade Federal do Rio de Janeiro (UFRJ), 21941-972 Rio de Janeiro, Rio de Janeiro, Brazil,
\email{fpinheiro@if.ufrj.br}
\and
R. Bachelard \at
	Departamento de F\'isica, Universidade Federal de S\~ao Carlos (UFSCar), 13565-905 S\~ao Carlos, S\~ao Paulo, Brazil,
\email{bachelard.romain@gmail.com}
\and
S. Slama \at
	Physikalisches Institut, Eberhardt-Karls-Universit\"{a}t T\"{u}bingen, D-72076 T\"{u}bingen, Germany,
\email{sebastian.slama@uni-tuebingen.de}
\and
Ph. W. Courteille \at				
	Instituto de F\'isica de S\~ao Carlos (IFSC), Universidade de S\~ao Paulo (USP), 13566-590 S\~ao Carlos, S\~ao Paulo, Brazil,
\email{philippe.courteille@ifsc.usp.br}
}
\maketitle

\abstract{
Despite a long history, light scattering by particles with size comparable with the light wavelength still unveils surprising optical phenomena, and many of them are related to the Fano effect.
Originally described in the context of atomic physics, the Fano resonance in light scattering arises from the interference between a narrow subradiant mode and a spectrally broad radiation line.
Here, we present an overview of Fano resonances in coated spherical scatterers within the framework of the Lorenz-Mie theory.
We briefly introduce the concept of conventional and unconventional Fano resonances in light scattering.
These resonances are associated with the interference between electromagnetic modes excited in the particle with different or the same multipole moment, respectively.
In addition, we investigate the modification of the spontaneous-emission rate of an optical emitter at the presence of a plasmonic nanoshell.
This modification of decay rate due to electromagnetic environment is referred to as the Purcell effect.
We analytically show that the Purcell factor related to a dipole emitter oriented orthogonal or tangential to the spherical surface can exhibit Fano or Lorentzian line shapes in the near field, respectively.
}

%\abstract*{Each chapter should be preceded by an abstract (10--15 lines long) that summarizes the content. The abstract will appear \textit{online} at \url{www.SpringerLink.com} and be available with unrestricted access. This allows unregistered users to read the abstract as a teaser for the complete chapter. As a general rule the abstracts will not appear in the printed version of your book unless it is the style of your particular book or that of the series to which your book belongs.\newline\indent
%Please use the 'starred' version of the new Springer \texttt{abstract} command for typesetting the text of the online abstracts (cf. source file of this chapter template \texttt{abstract}) and include them with the source files of your manuscript. Use the plain \texttt{abstract} command if the abstract is also to appear in the printed version of the book.}

%\tableofcontents

\section{Introduction}
\label{sec1:intro}

The Fano resonance, discovered in the realm of atomic physics by U. Fano in 1961~\cite{Fano_PhysRev124_1961}, is one of the hallmarks of interference in open quantum systems.
This interference effect was originally conceived as an interference between a transition to a bound state, coupled weakly to a continuum, and a transition directly to the same continuum~\cite{Fano_PhysRev124_1961}.
As a signature of quantum interference, the Fano effect has been extensively investigated in electronic transport at the nanoscale, in systems such as quantum dots, quantum wires, and tunnel junctions~\cite{Kivshar_RevModPhys82_2010}.

Being a wave interference phenomenon, Fano resonances are also present in classical optics and mechanics, where it can be understood as weak coupling between two classical oscillators driven by an external harmonic force~\cite{Kivshar_RevModPhys82_2010,Nussenzveig_AmJPhys70_2002}.
With the advent of metamaterials and plasmonic nanostructures, the Fano effect has recently become an important tool for tailoring and controlling electromagnetic mode interactions at subwavelength scale~\cite{Slama_NatPhys10_2014,Luk_NatMat9_2010}.
In plasmonics, it generally arises from the interference between a localized surface plasmon resonance and a spectrally broad superradiant mode acting as a background radiation~\cite{Kivshar_RevModPhys82_2010}.
Due to the sharpness of the Fano asymmetric line shape, systems exhibiting the Fano effect are highly sensitive to the local dielectric environment.
As a consequence, in plasmonic systems the Fano effect has been explored in the development of optical sensors, nonlinear devices, and low-threshold nanoscopic lasers~\cite{Luk_NatMat9_2010}.

Within the Lorenz-Mie scattering theory, the Fano effect results from the interference between electromagnetic modes excited in the scatterer with multipole moments of different orders (e.g., dipole-quadrupole interference)~\cite{Luk_NatMat9_2010} or same orders (e.g., dipole-dipole interference), which is sometimes referred to as unconventional Fano resonance~\cite{Tribelsky_EurophyLett97_2012,Arruda_PhysRevA87_2013,Gao_OptExp21_2013}.
In contrast to the conventional Fano resonance~\cite{Kivshar_JOpt15_2013}, the unconventional Fano effect in light scattering does not depend on the scattering direction, and it can be realized, e.g., with layered~\cite{Arruda_PhysRevA87_2013,Gao_OptExp21_2013,Monticone_PhysRevLett110_2013,Zayats_OptExp21_2013,Jelovina_Nano6_2014} or high-index~\cite{Tribelsky_EurophyLett97_2012,Limonov_OptExpress21_2013,Arruda_PhysRevA92_2015,Tribelsky_PhysRevA93_2016} particles.

Here, we study the influence of an unconventional Fano resonance of a plasmonic nanoshell on a single optical emitter in its vicinity~\cite{Arruda_PhysRevA96_2017}.  
The presence of a nanostructure is known to enhance the spontaneous-emission rate of optical emitters, which is generally referred to as the Purcell effect~\cite{Kerker_AppOpt19_1980,Chew_JCPhys87_1987,Klimov_OptComm211_2002,Vidal_PhysRevLett112_2014,Zadkov_PhysRevA90_2014}.
Many theoretical and experimental approaches have been developed to maximize~\cite{Datsyuk_PhysRevA75_2007,Szilar_PhysRevB94_2016,Liu_NatNano9_2014} or minimize~\cite{Farina_PhysRevA87_2013,Mahdifar_PhysRevA94_2016} the spontaneous-emission rate by changing the electromagnetic environment with engineered nanostructures.
In this chapter, we are interested in describing the connection between the Fano resonance usually observed in the Purcell factor~\cite{Kivshar_SciRep5_2015} and the unconventional Fano resonance exhibited by plasmonic nanoshells in light scattering~\cite{Arruda_PhysRevA87_2013,Arruda_PhysRevA92_2015,Arruda_PhysRevA96_2017}.

This chapter is organized as follows.
We recall the main analytical expressions of the Lorenz-Mie theory for light scattering by coated spherical particles in Sec.~\ref{sec2:Mie}.
The concept of conventional and unconventional Fano resonances in plasmonic nanoshells are briefly introduced.
In Sec.~\ref{sec3:Purcell}, we study the decay rates of single dipole emitters in the vicinity of plasmonic nanoshells.
Analytical expressions connecting Fano resonances in light scattering and the Purcell factor of dipole emitters are derived.
Finally, in Sec.~\ref{sec4:Conclusion}, we summarize our main results and contents of this chapter. 

\section{Light scattering by core-shell spheres: conventional and unconventional Fano resonances}
\label{sec2:Mie}

Light scattering by small particles is a fundamental topic in classical electrodynamics that has been studied and treated by several researchers, with applications ranging from meteorology and astronomy to biology and medicine~\cite{Bohren_Book_1983}. 
A complete analytic solution for homogeneous dielectric spheres with arbitrary radius was first derived, in an independent way, by L.V. Lorenz~\cite{Lorenz_KDan6_1890} and G. Mie~\cite{Mie_AnnPhys25_1908} more than a century ago. 
This solution, which is widely known as the Lorenz-Mie theory, is based on the expansion of the electromagnetic fields in terms of spherical wave functions~\cite{Bohren_Book_1983}.
An interesting generalization of this theory is the case of a spherical scatterer composed of materials with different optical properties, with the core-shell geometry being the simplest one.
Historically, the standard Lorenz-Mie theory, which deals with homogeneous spheres, was extended to single-layered spheres by Aden and Kerker~\cite{Kerker_JAppPhys22_1951} in 1951.
With the advent of plasmonics and metamaterials, core-shell systems have been extensively applied for experimental and theoretical investigations, such as the plasmonic cloaking technique~\cite{Alu_PhysRevE72_2005,Limonov_SciRep5_2015}, comb-like scattering response~\cite{Monticone_PhysRevLett110_2013}, tunable light scattering~\cite{Kivshar_OptLett38_2013,Arruda_PhysRevA94_2016}, fluorescence enhancement of optical emitters~\cite{Arruda_PhysRevA96_2017}, and Fano resonances~\cite{Halas_NanoLett10_2010}.
Indeed, the presence of cavities or dielectric materials inside metal-based nanostructures strongly modifies the scattering response due to the so-called plasmon hybridization~\cite{Nordlander_Science17_2003}.

In this section, we briefly recall the main analytical expressions used in the Lorenz-Mie theory for single-layered spheres.
Our aim is to introduce the concept of the Fano resonance in light scattering by plasmonic nanoshells, which will be further applied to the spontaneous-emission rate of single dipole emitters in Sec.~\ref{sec3:Purcell}.
With this aim, we present the complete theoretical framework in Sec.~\ref{subsec2:cross-sections}.
The discussion on plasmonic Fano resonances is treated in Sec.~\ref{subsec2:Fano} for a coated nanosphere composed of a silicon (Si) core and a silver (Ag) nanoshell. 
 
\subsection{The Lorenz-Mie theory for single-layered spheres}
\label{subsec2:cross-sections}

\begin{figure}[htbp]
\centering
\includegraphics[width=.9\textwidth]{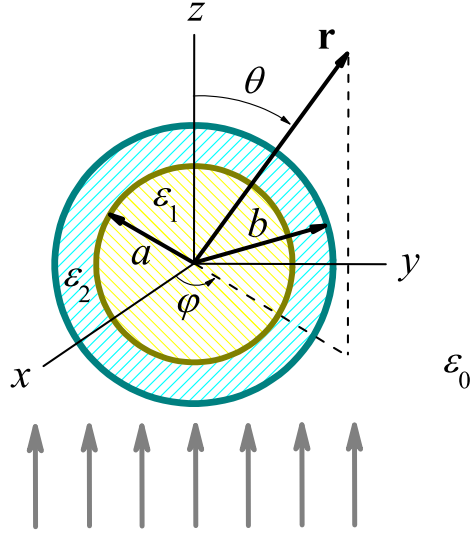}
\caption{A non-magnetic core-shell sphere interacting with an electromagnetic plane wave.
The inner sphere has radius $a$ and electric permittivity $\varepsilon_1$, whereas the outer sphere has radius $b$ and electric permittivity $\varepsilon_2$.
The surrounding medium is the vacuum $\varepsilon_0$. 
An electromagnetic plane wave propagating along the $z$ axis impinges on the sphere from below.}\label{fig1}
\end{figure}

Let us consider a coated sphere interacting with a plane wave $[\mathbf{E}(\mathbf{r}),\mathbf{H}(\mathbf{r})]e^{-\imath\omega t}$, where $\omega$ is the angular frequency.
The coated sphere is composed of a spherical core with radius $a$ and a single, center-symmetric shell with radius $b$, as depicted in Fig.~\ref{fig1}.
The involved media are assumed to be linear, homogeneous and isotropic.
In this case, the optical properties of media are described by a scalar electric permittivity $\varepsilon_p$ and a magnetic permeability $\mu_p$, with label $p=1$ for the core $(0\leq r\leq a)$, $p=2$ for the shell $(a\leq r\leq b)$ and $p=0$ for the surrounding medium $(r\geq b)$, which is assumed to be the vacuum. 
At optical frequencies, naturally occurring media are usually non-magnetic: $\mu_1=\mu_2=\mu_0$.

The macroscopic Maxwell's equations associated with the system illustrated in Fig.~\ref{fig1} provide the vector Helmholtz equation $(\nabla^2+k^2)[\mathbf{E}(\mathbf{r}),\mathbf{H}(\mathbf{r})]=(\mathbf{0},\mathbf{0})$, where $k=2\pi/\lambda$ is the wave number and $\lambda$ is the wavelength of the light in each medium $p=\{0,1,2\}$.
The interested reader is refereed to Ref.~\cite{Bohren_Book_1983} for a complete and detailed solution of this vector equation.
Since the sphere material is non-optically active~\cite{Arruda_JOSA30_2013}, without loss of generality, we consider the polarization of the incident wave along the $x$-direction.
In terms of spherical wave functions, the incident and scattered electric fields ($r\geq b$) can be cast as
\begin{eqnarray}
\mathbf{E}_{\rm in}(r,\theta,\varphi)&=&-\frac{1}{kr}\sum_{\ell=1}^{\infty}E_{\ell}\bigg\{\imath\cos\varphi\sin\theta j_{\ell}(kr)\ell(\ell+1)\pi_{\ell}(\cos\theta)\hat{\bf e}_r\nonumber\\
&&-\cos\varphi\left[\pi_{\ell}(\cos\theta)\psi_{\ell}(kr)-\imath\tau_{\ell}(\cos\theta)\psi_{\ell}'(kr)\right]\hat{\bf e}_{\theta}\nonumber\\
&&-\sin\varphi\left[\imath\pi_{\ell}(\cos\theta)\psi_{\ell}'(kr)-\tau_{\ell}(\cos\theta)\psi_{\ell}(kr)\right]\hat{\bf e}_{\varphi}\bigg\},\label{Ein}\\
\mathbf{E}_{\rm sca}(r,\theta,\varphi)&=&\frac{1}{kr}\sum_{\ell=1}^{\infty}E_{\ell}\bigg\{\imath\cos\varphi\sin\theta a_{\ell} h_{\ell}^{(1)}(kr)\ell(\ell+1)\pi_{\ell}(\cos\theta)\hat{\bf e}_r\nonumber\\
&&-\cos\varphi\left[b_{\ell}\pi_{\ell}(\cos\theta)\xi_{\ell}(kr)-\imath a_{\ell}\tau_{\ell}(\cos\theta)\xi_{\ell}'(kr)\right]\hat{\bf e}_{\theta}\nonumber\\
&&-\sin\varphi\left[\imath a_{\ell}\pi_{\ell}(\cos\theta)\xi_{\ell}'(kr)-b_{\ell}\tau_{\ell}(\cos\theta)\xi_{\ell}(kr)\right]\hat{\bf e}_{\varphi}\bigg\},\label{Esca}
\end{eqnarray}
where $k=\omega\sqrt{\varepsilon_0\mu_0}$, $E_{\ell} = \imath^{\ell} E_0(2\ell+1)/[\ell(\ell+1)]$, $\pi_{\ell}(\cos\theta)=P_{\ell}^1(\cos\theta)/\sin\theta$, $\tau_{\ell}(\cos\theta)={\rm d}P_{\ell}^1(\cos\theta)/{\rm d}\theta$, with $P_{\ell}^1$ being the associated Legendre function of first order.
The coefficients $a_{\ell}$ and $b_{\ell}$ are the transverse magnetic (TM) and transverse electric (TE) Lorenz-Mie coefficients, respectively, and are determined from boundary conditions.
For center-symmetric coated spheres, these coefficients read~\cite{Bohren_Book_1983,Arruda_JOpt14_2012}:
\begin{eqnarray}
        a_{\ell} &=&\frac{\widetilde{n}_2\psi_{\ell}'(kb)-\psi_{\ell}(kb)\mathcal{A}_{\ell}(n_2kb)}{\widetilde{n}_2\xi_{\ell}'(kb)-\xi_{\ell}(kb)\mathcal{A}_{\ell}(n_2kb)}\
        ,\label{an}\\
        b_{\ell} &=&\frac{\psi_{\ell}'(kb)-\widetilde{n}_2\psi_{\ell}(kb)\mathcal{B}_{\ell}(n_2kb)}{\xi_{\ell}'(kb)-\widetilde{n}_2\xi_{\ell}(kb)\mathcal{B}_{\ell}(n_2kb)}\
        ,\label{bn}
    \end{eqnarray}
with the auxiliary functions
\begin{eqnarray}
\mathcal{A}_{\ell}(n_2kb)&=&\frac{\psi_{\ell}'(n_2kb)-A_{\ell}\chi_{\ell}'(n_2kb)}{\psi_{\ell}(n_2kb)-A_{\ell}\chi_{\ell}(n_2kb)},\\
\mathcal{B}_{\ell}(n_2kb)&=&\frac{\psi_{\ell}'(n_2kb)-B_{\ell}\chi_{\ell}'(n_2kb)}{\psi_{\ell}(n_2kb)-B_{\ell}\chi_{\ell}(n_2kb)},\\
        A_{\ell} &=& \frac{\widetilde{n}_2\psi_{\ell}(n_2ka)\psi_{\ell}'(n_1ka)-\widetilde{n}_1\psi_{\ell}'(n_2ka)\psi_{\ell}(n_1ka)}{\widetilde{n}_2\chi_{\ell}(n_2ka)\psi_{\ell}'(n_1ka)-\widetilde{n}_1\chi_{\ell}'(n_2ka)\psi_{\ell}(n_1ka)},\label{An}\\
        B_{\ell}&=&\frac{\widetilde{n}_2\psi_{\ell}'(n_2ka)\psi_{\ell}(n_1ka)-\widetilde{n}_1\psi_{\ell}(n_2ka)\psi_{\ell}'(n_1ka)}{\widetilde{n}_2\chi_{\ell}'(n_2ka)\psi_{\ell}(n_1ka)-\widetilde{n}_1\chi_{\ell}(n_2ka)\psi_{\ell}'(n_1ka)},\label{Bn}
\end{eqnarray}
where the functions $\psi_{\ell}(z)=z j_{\ell}(z)$, $\chi_{\ell}(z)=-z y_{\ell}(z)$ and $\xi_{\ell}(z)=\psi_{\ell}(z)-\imath\chi_{\ell}(z)$ are the Riccati-Bessel, Riccati-Neumann and Riccati-Hankel functions, respectively, with $j_{\ell}$ and $y_{\ell}$ being the spherical Bessel and Neumann functions~\cite{Bohren_Book_1983}.
The refractive and impedance indices are $n_p=\sqrt{\varepsilon_{p}\mu_{p}/(\varepsilon_0\mu_0)}$ and $\widetilde{n}_p = \sqrt{\varepsilon_{p}\mu_0/(\varepsilon_0\mu_{p})}$, with $p=\{1,2\}$~\cite{Arruda_JOpt14_2012}.
For non-magnetic materials ($\mu_p=\mu_0$), one has $\widetilde{n}_p=n_p$~\cite{Arruda_JOSA27_1_2010}.
The solution for a homogeneous sphere of radius $b$ can be readily obtained by setting $\varepsilon_1=\varepsilon_2$ and $\mu_1=\mu_2$, i.e., $A_{\ell}=0=B_{\ell}$.
It is worth mentioning that these Lorenz-Mie coefficients can be trivially generalized to the case of center-symmetric multilayered spheres~\cite{Wang_RadSci26_1991}.

Analogously, within the core $(0\leq r\leq a)$ and shell $(a\leq r\leq b)$ regions, we have the electric fields~\cite{Arruda_JOpt14_2012,Kaiser_AppOpt33_1994}
\begin{eqnarray}
\mathbf{E}_{1}(r,\theta,\varphi)&=&-\frac{1}{n_1kr}\sum_{\ell=1}^{\infty}E_{\ell}\bigg\{\imath\cos\varphi\sin\theta d_{\ell} j_{\ell}(n_1kr)\ell(\ell+1)\pi_{\ell}(\cos\theta)\hat{\bf e}_r\nonumber\\
&&+\cos\varphi\left[c_{\ell}\pi_{\ell}(\cos\theta)\psi_{\ell}(n_1kr)-\imath d_{\ell}\tau_{\ell}(\cos\theta)\psi_{\ell}'(n_1kr)\right]\hat{\bf e}_{\theta}\nonumber\\
&&+\sin\varphi\left[\imath d_{\ell}\pi_{\ell}(\cos\theta)\psi_{\ell}'(n_1kr)-c_{\ell}\tau_{\ell}(\cos\theta)\psi_{\ell}(n_1kr)\right]\hat{\bf e}_{\varphi}\bigg\},\label{E1}\\
\mathbf{E}_{2}(r,\theta,\varphi)&=&-\frac{1}{n_2kr}\sum_{\ell=1}^{\infty}E_{\ell}\bigg\{\imath\cos\varphi\sin\theta g_{\ell} j_{\ell}(n_2kr)\ell(\ell+1)\pi_{\ell}(\cos\theta)\hat{\bf e}_r\nonumber\\
&&+\imath\cos\varphi\sin\theta w_{\ell} y_{\ell}(n_2kr)\ell(\ell+1)\pi_{\ell}(\cos\theta)\hat{\bf e}_r\nonumber\\
&&+\cos\varphi\left[f_{\ell}\pi_{\ell}(\cos\theta)\psi_{\ell}(n_2kr)-\imath g_{\ell}\tau_{\ell}(\cos\theta)\psi_{\ell}'(n_2kr)\right]\hat{\bf e}_{\theta}\nonumber\\
&&-\cos\varphi\left[v_{\ell}\pi_{\ell}(\cos\theta)\chi_{\ell}(n_2kr)-\imath w_{\ell}\tau_{\ell}(\cos\theta)\chi_{\ell}'(n_2kr)\right]\hat{\bf e}_{\theta}\nonumber\\
&&+\sin\varphi\left[\imath g_{\ell}\pi_{\ell}(\cos\theta)\psi_{\ell}'(n_2kr)-f_{\ell}\tau_{\ell}(\cos\theta)\psi_{\ell}(n_2kr)\right]\hat{\bf e}_{\varphi}\nonumber\\
&&-\sin\varphi\left[\imath w_{\ell}\pi_{\ell}(\cos\theta)\chi_{\ell}'(n_2kr)-v_{\ell}\tau_{\ell}(\cos\theta)\chi_{\ell}(n_2kr)\right]\hat{\bf e}_{\varphi}\bigg\},\label{E2}
\end{eqnarray}
respectively.
In terms of the auxiliary functions defined in Eqs.~(\ref{An}) and (\ref{Bn}), the Lorenz-Mie coefficients $c_{\ell}$, $d_{\ell}$, $f_{\ell}$, $g_{\ell}$, $v_{\ell}$ and $w_{\ell}$ read~\cite{Bohren_Book_1983,Arruda_JOpt14_2012}
    \begin{eqnarray}
        c_{\ell} &=&\frac{n_1f_{\ell}}{n_2\psi_{\ell}(n_1ka)}\left[\psi_{\ell}(n_2ka)-B_{\ell}\chi_{\ell}(n_2ka)\right],\\
        d_{\ell} &=&\frac{n_1g_{\ell}}{n_2\psi_{\ell}'(n_1ka)}\left[\psi_{\ell}'(n_2ka)-A_{\ell}\chi_{\ell}'(n_2ka)\right],\\
    f_{\ell}&=&\frac{\imath
    n_2}{\left[\psi_{\ell}(n_2kb)-B_{\ell}\chi_{\ell}(n_2kb)\right]\left[\xi_{\ell}'(kb)-\widetilde{n}_2\xi_{\ell}(kb)\mathcal{B}_{\ell}(n_2kb)\right]},\\
    g_{\ell}&=&\frac{\imath
    n_2}{\left[\psi_{\ell}(n_2kb)-A_{\ell}\chi_{\ell}(n_2kb)\right]\left[\widetilde{n}_2\xi_{\ell}'(kb)-\xi_{\ell}(kb)\mathcal{A}_{\ell}(n_2kb)\right]},\label{gn} \\
        v_{\ell} &=&B_{\ell}f_{\ell} ,\\
        w_{\ell} &=&A_{\ell}g_{\ell} .\label{wn}
    \end{eqnarray}

Equations (\ref{Ein})--(\ref{wn}) are the complete Lorenz-Mie solution for center symmetric core-shell spheres~\cite{Bohren_Book_1983}.
The corresponding magnetic field $\mathbf{H}(\mathbf{r})$ can be straightforwardly obtained from Eqs.~(\ref{Ein}), (\ref{Esca}), (\ref{E1}), and (\ref{E2}) by Maxwell's curl equations. 
In the following, we discuss the cross sections and internal field intensities in the context of Fano resonances in plasmonic nanoshells.

\subsection{Fano resonances in optical cross sections}
\label{subsec2:Fano}

The cross sections of a spherical particle can be calculated exactly from the net rate of electromagnetic energy crossing an imaginary surface at the far field (for details, see Ref.~\cite{Bohren_Book_1983}).
From the standard Lorenz-Mie theory, by using Eqs.~(\ref{Ein}) and (\ref{Esca}), the extinction, scattering and absorption cross sections of a spherical particle irradiated by plane waves are, respectively,
\begin{eqnarray}
\sigma_{\rm ext} &=& \frac{2\pi}{k^2}\sum_{\ell=1}^{\infty}(2\ell+1){\rm Re}\left(a_{\ell}+b_{\ell}\right),\label{Qext}\\
\sigma_{\rm sca} &=& \frac{2\pi}{k^2}\sum_{\ell=1}^{\infty}(2\ell+1)\left(|a_{\ell}|^2+|b_{\ell}|^2\right),\\
\sigma_{\rm abs} &=& \sigma_{\rm ext}-\sigma_{\rm sca},\label{Qabs}
\end{eqnarray}
where $a_{\ell}$ and $b_{\ell}$ carry the dependence on the geometrical and optical parameters of the scatterer, and are defined in Eqs.~(\ref{an}) and (\ref{bn}) for a single-layered core-shell sphere.
Equations (\ref{Qext})--(\ref{Qabs}) are calculated from averaging over all possible directions and polarizations.
By considering the backward ($\theta=\pi$) and forward $(\theta=0)$ directions, we obtain 
\begin{eqnarray}
\sigma_{\rm back} &=& \frac{\pi}{k^2}\left|\sum_{\ell=1}^{\infty}(2\ell+1)(-1)^{\ell}\left(a_{\ell}-b_{\ell}\right)\right|^2,\label{Qback}\\
\sigma_{\rm forward} &=& \frac{\pi}{k^2}\left|\sum_{\ell=1}^{\infty}(2\ell+1)\left(a_{\ell}+b_{\ell}\right)\right|^2,\label{Qforward}
\end{eqnarray}
which are the differential backward and forward scattering cross sections, respectively.
Usually the optical cross sections are calculated in units of the geometrical cross section $\sigma_{\rm g}=\pi b^2$, where $b$ is the effective radius of the spherical scatterer.

From Eqs.~(\ref{Qext})--(\ref{Qforward}), it is clear that one can achieve interferences between different electric and magnetic scattering amplitudes (namely, $a_{\ell}$ and $b_{\ell}$) only for directional scattering, e.g., $\sigma_{\rm back}$ and $\sigma_{\rm forward}$~\cite{Kivshar_JOpt15_2013}.
Of particular interest is the case of light scattering by small plasmonic spheres $(kb\leq1)$.
In this limiting case, the dipolar Rayleigh scattering $(\ell=1)$ plays the role of a broad spectral resonance, whereas the localized surface plasmon resonance, e.g., quadrupole $(\ell=2)$ or higher order resonance, plays the role of a narrow spectral line interacting with a broad spectral line.
As a result, in the vicinity of the narrow plasmon resonance there is a $\pi$-phase jump, leading to the coexistence of constructive and destructive interferences with the broad dipole resonance.
This interference between the electric scattering amplitudes $a_1$ and $a_2$ is described by a characteristic asymmetric line shape, known as the conventional Fano resonance.

\subsubsection{Unconventional Fano resonances in plasmonic nanoshells}

Recently, other mechanisms of Fano-like resonances have been described in light scattering by small particles relative to the light wavelength.
For instance, Fano resonances were shown to occur beyond the applicability of the Rayleigh approximation in high-index particles, where the interference between electromagnetic modes with the same multipole moment (e.g., dipole-dipole interference) is crucial~\cite{Tribelsky_EurophyLett97_2012,Limonov_OptExpress21_2013,Tribelsky_PhysRevA93_2016}.
These Fano-like resonances also manifest themselves in plasmonic layered particles with moderate permittivities~\cite{Monticone_PhysRevLett110_2013}, even in the Rayleigh scattering approximation~\cite{Arruda_PhysRevA87_2013}.
Since these interferences occur in the total scattering cross section $\sigma_{\rm sca}$ and, hence, do not depend on the scattering direction, they were named unconventional Fano resonances~\cite{Tribelsky_EurophyLett97_2012}.

To picture these concepts, let us consider a core-shell nanoparticle consisting of a silicon (Si) core with refractive index $n_1=3.5$ and radius $a=60$~nm coated with a dispersive silver (Ag) nanoshell with radius $b=90$~nm.
The Ag dielectric permittivity is well described by the generalized Drude model~\cite{Christy_PhysRevB6_1972,Boltasseva_LaserPhotRev4_2010} 
\begin{equation}
\frac{\varepsilon_{\rm Ag}(\omega)}{\varepsilon_0}=\varepsilon_{\rm int}-\frac{\omega_{\rm p}^2}{\omega(\omega+\imath\gamma)},\label{eps-Ag}
\end{equation}
where $\varepsilon_{\rm int}=3.7$ is a contribution due to interband transitions, $\omega_{\rm p}=9.2$~eV ($\approx 2\pi\times2.2\times10^{15}$~Hz) is the plasmon frequency associated with conduction electrons, and $\gamma=0.02$~eV is the effective dumping rate due to material losses.
These Drude parameters for Ag are valid below the frequency of onset for interband transitions: $\omega/\omega_{\rm p}<0.42$~\cite{Boltasseva_LaserPhotRev4_2010}.
These are the optical and geometric parameters that we consider for numerical calculations throughout this chapter.

\begin{figure}[htbp]
\centering
\includegraphics[width=\textwidth]{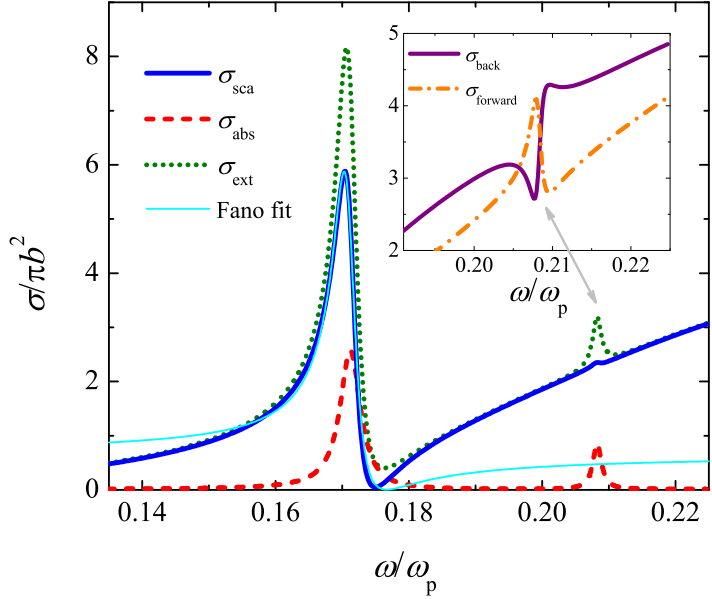}
\caption{Optical cross sections in the light scattering by a (Si) core-shell (Ag) nanosphere in free space.
The dielectric core has radius $a=60$~nm and refractive index $n_1=3.5$, whereas the plasmonic shell has radius $b=90$~nm and electric permittivity $\varepsilon_2=\varepsilon_{\rm Ag}(\omega)$ [Eq.~(\ref{eps-Ag})].
The plot shows the scattering ($\sigma_{\rm sca}$), absorption ($\sigma_{\rm abs}$), and extinction ($\sigma_{\rm ext}$) cross sections (in units of $\pi b^2$) as a function of the frequency $\omega$ (in units of Ag plasmon frequency $\omega_{\rm p}$).
An unconventional Fano resonance can be observed in $\sigma_{\rm sca}$ ($\omega\approx 0.170\omega_{\rm p}$) associated with the dipole-dipole interference $a_1a_1^*$ excited in the shell, where $a_1$ is the electric Lorenz-Mie coefficient.
The inset shows two conventional Fano resonances in the differential backward $(\sigma_{\rm back})$ and forward $(\sigma_{\rm forward}$) scattering cross sections at $\omega\approx 0.208\omega_{\rm p}$.
These Fano resonances are related to the dipole-quadrupole interference $a_1a_2^*$ at the backward and forward directions, respectively.}\label{fig2}
\end{figure}

Figure~\ref{fig2} shows the plots of the optical cross sections defined in Eqs.~(\ref{Qext})--(\ref{Qforward}) as a function of the frequency of the incident electromagnetic wave.
For the frequency range $0.135\omega_{\rm p}<\omega<0.225\omega_{\rm p}$, the corresponding size parameters of the core-shell sphere are $0.56<kb<0.95$, so we can restrict our discussion on electric multipole moments up to $\ell=2$ (quadrupole).
Also, since the involved materials are non-magnetic with moderate permittivities, one has $b_{\ell}\approx 0$ for $kb<1$.

In the main plot of Fig.~\ref{fig2}, one can clearly see that $\sigma_{\rm sca}$ presents a Fano line shape, where the dipole-dipole $(|a_1|^2)$ resonance occurs at $\omega\approx 0.170\omega_{\rm p}$ and the antiresonance (Fano dip) occurs at $\omega\approx 0.175\omega_{\rm p}$.
In this same frequency range, the absorption cross section $\sigma_{\rm abs}$ exhibits a Lorentzian line shape~\cite{Ruan_PhysChemC114_2010}.
In addition, a quadrupole-quadrupole $(|a_2|^2)$ resonance also shows up at $\omega\approx0.208\omega_{\rm p}$, but only contributes to the absorption cross section.
However, the overlap of the narrow quadrupole ($\ell=2$) resonance and the broad dipole resonance ($\ell=1$) leads to a Fano line shape in the differential scattering cross sections, see the inset of Fig.~\ref{fig2}.

The unconventional Fano resonance observed in $\sigma_{\rm sca}$ and $\sigma_{\rm ext}$ can be explained by the interference between out of phase electric fields within the plasmonic nanoshell.
Recently, Tribelsky and Miroshnichenko~\cite{Tribelsky_PhysRevA93_2016} have shown that the Fano line shape associated with high-index spherical particles can be calculated exactly within the Lorenz-Mie theory.
Here, we generalize their result to the case of a core-shell sphere.
Since we are not interested in magnetic resonances (namely, $b_{\ell}$)~\cite{Arruda_JOSA27_1_2010,Arruda_JOSA27_2_2010}, we restrict our discussion on the electric scattering amplitude $a_{\ell}$.
Indeed, the magnetic case is completely analogous and the interested reader is referred to Ref.~\cite{Tribelsky_PhysRevA93_2016}. 

Following Ref.~\cite{Tribelsky_PhysRevA93_2016}, we rewrite the electric scattering coefficient $a_{\ell}$:
\begin{eqnarray}
a_{\ell} = \frac{{F}_{\ell}}{{F}_{\ell}+\imath {G}_{\ell}} =\frac{\zeta_{\ell}(\omega) + q_{\ell}}{\zeta_{\ell}(\omega) + q_{\ell} - \imath\left[\zeta_{\ell}(\omega)q_{\ell}-1\right]},\label{a1}
\end{eqnarray}
with the new auxiliary functions being 
\begin{eqnarray}
{F}_{\ell} &=& n_2\psi_{\ell}'(kb)\left[\psi_{\ell}(n_2kb)-A_{\ell}\chi_{\ell}(n_2kb)\right]\nonumber\\
&&-\psi_{\ell}(kb)\left[\psi_{\ell}'(n_2kb)-A_{\ell}\chi_{\ell}'(n_2kb)\right],\\
{G}_{\ell} &=& -n_2\chi_{\ell}'(kb)\left[\psi_{\ell}(n_2kb)-A_{\ell}\chi_{\ell}(n_2kb)\right]\nonumber\\
&&+\chi_{\ell}(kb)\left[\psi_{\ell}'(n_2kb)-A_{\ell}\chi_{\ell}'(n_2kb)\right],
\end{eqnarray}
where $\zeta_{\ell}(\omega)\equiv\zeta_{\ell}'(\omega)+\imath\zeta_{\ell}''(\omega)$ and $q_{\ell}$ is the Fano asymmetry parameter.
Here, $\zeta'={\rm Re}(\zeta)$ and $\zeta''={\rm Im}(\zeta)$ (not to be confused with derivatives with respect to the argument).
Although the demonstration is not trivial~\cite{Tribelsky_PhysRevA93_2016}, one can formally show that
\begin{eqnarray}
\zeta_{\ell}(\omega) &=& \frac{{F}_{\ell}\psi_{\ell}'(kb) - {G}_{\ell}\chi_{\ell}'(kb)}{\psi_{\ell}'(n_2kb)-A_{\ell}\chi_{\ell}'(n_2kb)},\label{zeta}\\
q_{\ell}&=& \frac{\chi_{\ell}'(kb)}{\psi_{\ell}'(kb)}.\label{q-LM}
\end{eqnarray}
If the sphere is lossless, one has $\zeta_{\ell}''(\omega)=0$ and $|a_{\ell}|^2=(\zeta_{\ell}'+q_{\ell})^2/[(1+q_{\ell}^2)(\zeta_{\ell}'^2+1)]$, i.e., $|a_{\ell}|^2$ is a normalized Fano lineshape as a function of $\zeta_{\ell}'$.
These expressions agree with Ref.~\cite{Tribelsky_PhysRevA93_2016} for $A_{\ell}=0$ (homogeneous sphere).

Considering only the dipole scattering resonance ($\ell=1$) and defining $q_{\rm LM}\equiv q_1$ and $\zeta(\omega)\equiv\zeta_1(\omega)$, we finally have 
\begin{equation}
\sigma_{\rm sca}\approx\frac{6\pi}{k^2\left(1+q_{\rm LM}^2\right)}\left\{\displaystyle\frac{\left[\displaystyle\frac{\zeta'(\omega)}{1+\zeta''(\omega)}+\displaystyle\frac{q_{\rm LM}}{1+\zeta''(\omega)}\right]^2 + \left[\displaystyle\frac{\zeta''(\omega)}{1+\zeta''(\omega)}\right]^2}{\left[\displaystyle\frac{\zeta'(\omega)}{1+\zeta''(\omega)}\right]^2+1}\right\}.\label{sigma-approx}
\end{equation}
In the vicinity of a Fano resonance, one can use the approximation $\zeta_{\ell}'(\omega)/[1+\zeta''(\omega)]\approx (\omega - \omega_{\rm res})/\Omega$, where $\Omega$ is associated with the curve linewidth.
The function $\zeta''(\omega)$ has a very complicated analytical expression, and it can be estimated from the dipole resonance $[\zeta'(\omega_{\rm res})=0]$: $\sigma_{\rm sca}^{(\rm max)} =6\pi(q^2 + \zeta''^2)/[(k^2(1+q^2)(1+\zeta''^2)]$.
From Fig.~\ref{fig2}, one has $\sigma_{\rm sca}^{\rm max}(\omega_{\rm res})\approx 5.9\pi b^2$ for $\omega_{\rm res}\approx0.170\omega_{\rm p}$.
Indeed, we have used Eq.~(\ref{sigma-approx}) to fit the scattering cross section in Fig.~\ref{fig2}.
For our set of parameters, the effective Fano asymmetry parameter is $q_{\rm LM}/(1+\zeta'')\approx -2.81$, where $q_{\rm LM}\approx-3.84$ and $\zeta''\approx0.368$.

\subsubsection{Off-resonance field enhancement in plasmonic nanoshells}

\begin{figure*}[htbp]
{\includegraphics[width=.6\textwidth]{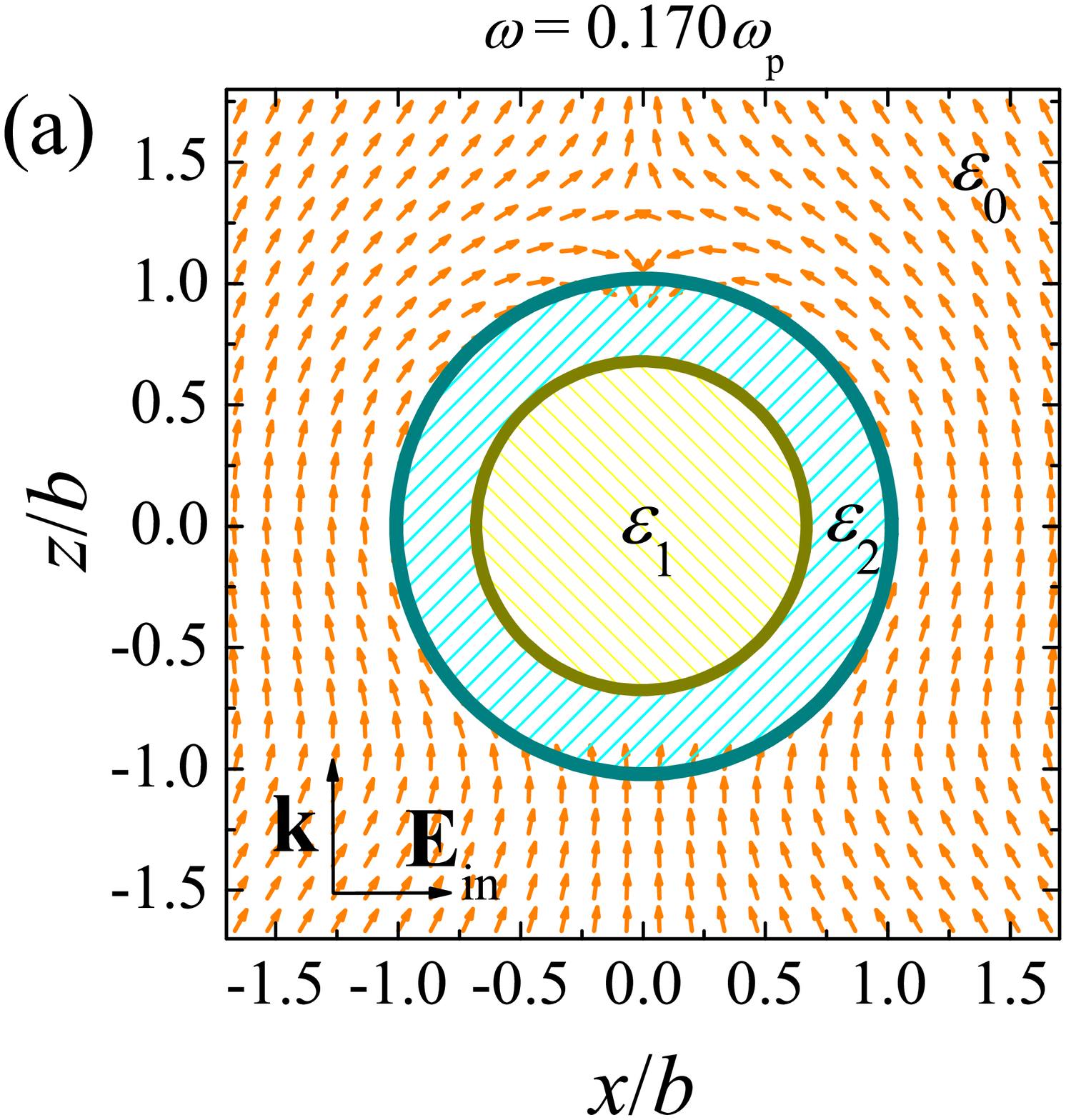}}\hspace{-1.6cm}
{\includegraphics[width=.6\textwidth]{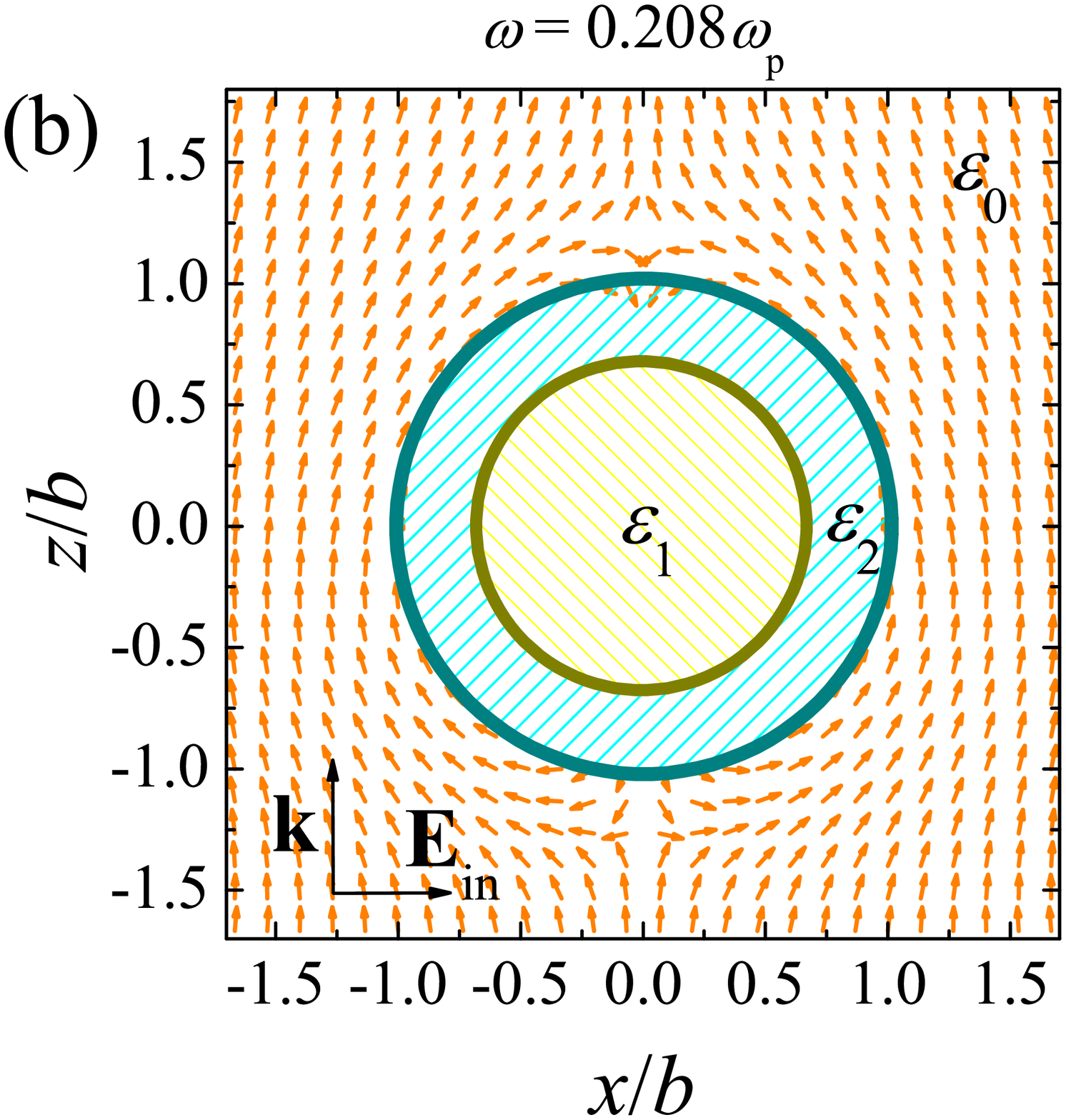}}
{\includegraphics[width=.6\textwidth]{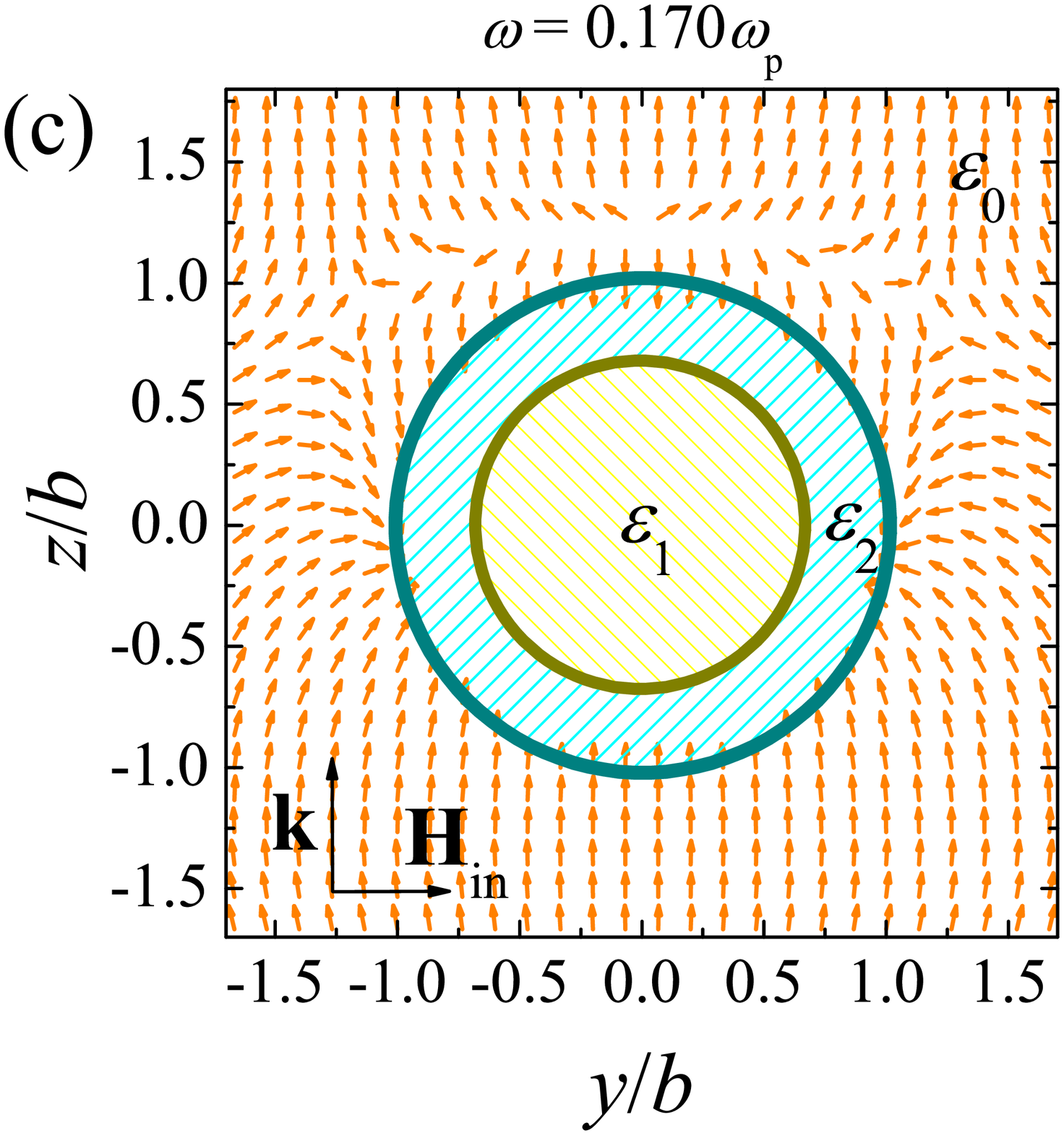}}\hspace{-1.6cm}
{\includegraphics[width=.6\textwidth]{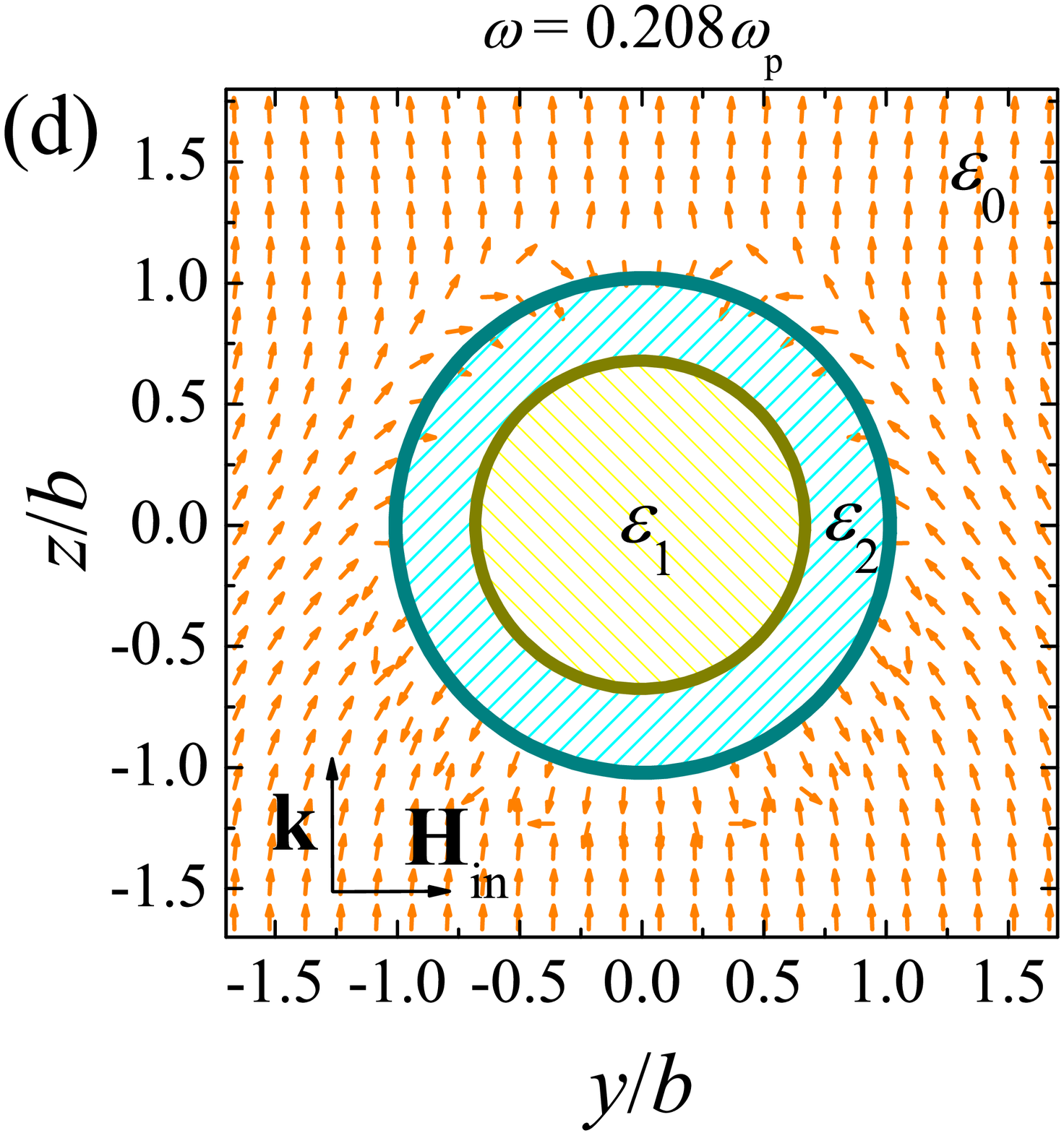}}
\caption{Time-averaged energy flow (normalized Poynting vector field) in the vicinity of a (Si) core-shell (Ag) nanosphere for dipole ($\omega\approx0.170\omega_{\rm p}$) and quadrupole ($\omega\approx0.208\omega_{\rm p}$) scattering resonances.
The dielectric core has refractive index $n_1=3.5$ and radius $a=60$~nm, whereas the Ag nanoshell [Eq.~(\ref{eps-Ag})] has radius $b=90$~nm.
The $xz$ plane shows the presence of a saddle point in the energy flow in the $z$-axis around $z\approx 1.25b$ for dipole resonance (a) and two saddle points for quadrupole resonance (b) around $z\approx\pm 1.20b$.
The $yz$ plane shows singular points along the $y$ direction for dipole (c) and quadrupole (d) resonances.}\label{fig3}
\end{figure*}

The presence of Fano-like resonances in Lorenz-Mie theory is associated with very interesting optical phenomena, such as the formation of optical vortices and saddle points in the energy flow around particles~\cite{Kivshar_JOpt15_2013}, enhanced light scattering response~\cite{Kivshar_OptLett38_2013}, and off-resonance field enhancement within core-shell scatterers~\cite{Miroshnichenko_PhysRevA81_2010,Arruda_JOSA32_2015}. 
Indeed, as can be observed in Fig.~\ref{fig3}, both dipole and quadrupole scattering resonances discussed above are associated with saddle points in the time-averaged energy flow $\mathbf{S}(\mathbf{r})={\rm Re}[\mathbf{E}(\mathbf{r})\times\mathbf{H}^*(\mathbf{r})]$/2 in the vicinity of the Ag nanoshell, where the local electromagnetic field $(\mathbf{E},\mathbf{H})$ is calculated from Eqs.~(\ref{Ein}) and (\ref{Esca}).

\begin{figure}[htbp]
\centering
\includegraphics[width=\textwidth]{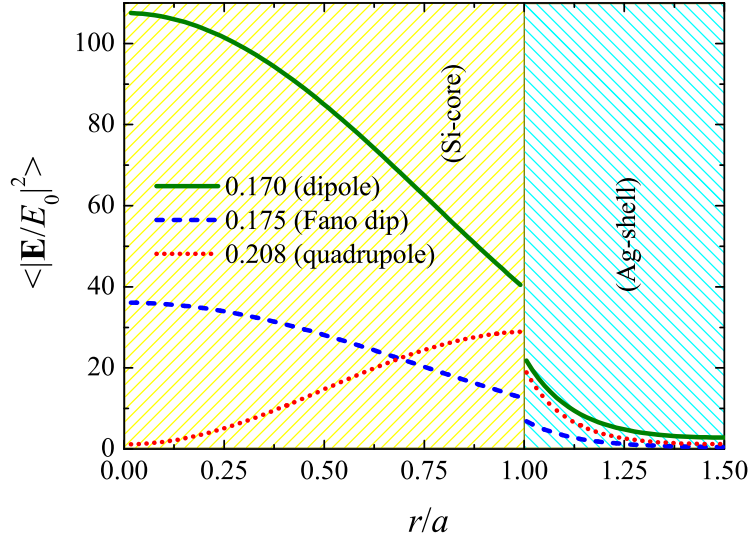}
\caption{The angle-averaged electric field intensity inside a (Si) core-shell (Ag) nanosphere in free space, as depicted in Fig.~\ref{fig1}, as a function of the distance from the center of the scatterer to its surface.
The dielectric core has radius $a=60$~nm and refractive index $n_1=3.5$, whereas the plasmonic shell has radius $b=90$~nm and electric permittivity $\varepsilon_2=\varepsilon_{\rm Ag}(\omega)$ [Eq.~(\ref{eps-Ag})].
The maximum electric field intensity stored inside the scatterer occurs at the dipole resonance ($\omega\approx 0.170\omega_{\rm p}$). 
The intensity at the Fano dip ($\omega\approx 0.175\omega_{\rm p}$) is comparable to and even greater than the intensity at the quadrupole resonance ($\omega\approx 0.208\omega_{\rm p}$) inside the core.}\label{fig4}
\end{figure}

To show the effect of off-resonance field enhancement, we calculate the corresponding electric field intensity $\langle |\mathbf{E}|^2\rangle$ within the coated sphere~\cite{Arruda_PhysRevA87_2013,Kaiser_AppOpt33_1994,Arruda_JOSAA34_2017}.
Here, the operator $\langle \cdots \rangle = (1/{4\pi})\int_{-1}^{1}{\rm d}(\cos\theta)\int_0^{2\pi}{\rm d}\varphi (\cdots)$ is the angle average over 4$\pi$.
Using the exact expression for the electric fields within the core $(0\leq r\leq a)$ and shell $(a\leq r\leq b)$, Eqs.~(\ref{E1}) and (\ref{E2}), we obtain the angle-averaged intensities~\cite{Arruda_PhysRevA87_2013,Kaiser_AppOpt33_1994}
\begin{eqnarray}
    \frac{\langle |\mathbf{E}_1(\mathbf{r})|^2\rangle}{|E_0|^2}&=&\frac{1}{2}\sum_{\ell=1}^{\infty}\bigg\{(2\ell+1) |c_{\ell} |^2 |j_{\ell}(n_1kr)|^2\nonumber \\
    &&+ |d_{\ell} |^2\left[\ell
    |j_{\ell+1}(n_1kr)|^2+(\ell+1)|j_{\ell-1}(n_1kr)|^2\right]\bigg\},\label{E1-med}\\
    \frac{\langle |\mathbf{E}_2(\mathbf{r})|^2\rangle}{|E_0|^2}&=&\frac{1}{2}\sum_{n=1}^{\infty}\Bigg\{(2\ell+1) \left[|f_{\ell} |^2 |j_{\ell}(n_2kr)|^2+  |v_{\ell} |^2 |y_{\ell}(n_2kr)|^2\right]\nonumber\\
    &&+ |g_{\ell} |^2 \left[\ell
    |j_{\ell+1}(n_2kr)|^2+(\ell+1)|j_{\ell-1}(n_2kr)|^2
    \right]\nonumber\\
    &&+ |w_{\ell} |^2 \left[\ell
    |y_{\ell+1}(n_2kr)|^2+(\ell+1)|y_{\ell-1}(n_2kr)|^2
    \right]\nonumber\\
    &&+ 2{\rm Re}\bigg[ (2\ell+1) f_{\ell} v_{\ell}^{*} j_{\ell}(n_2kr)y_{\ell}(n_2^*kr)\nonumber\\
    &&+ g_{\ell} w_{\ell}^{*}\big[\ell j_{\ell+1}(n_2kr)y_{\ell+1}(n_2^*kr)\nonumber\\
    &&+(\ell+1)j_{\ell-1}(n_2kr)y_{\ell-1}(n_2^*kr)
    \big]\bigg]\Bigg\},\label{E2-med}
\end{eqnarray}
where we have used the relations~\cite{Bohren_Book_1983}: $(2\ell+1)\int_{-1}^1{\rm d}(\cos\theta)(\pi_{\ell}\pi_{\ell'}+\tau_{\ell}\tau_{\ell'})=2\ell^2(\ell+1)^2\delta_{\ell\ell'}$, $\int_{-1}^1{\rm d}(\cos\theta)(\pi_{\ell}\tau_{\ell'}+\tau_{\ell}\pi_{\ell'})=0$, and $(2\ell+1)\int_{-1}^1{\rm d}(\cos\theta)\pi_{\ell}\pi_{\ell'}\sin^2\theta=2\ell(\ell+1)\delta_{\ell\ell'}$, with $\delta_{\ell\ell'}$ being the Kronecker delta.
Note that the electric field intensity $\langle |\mathbf{E}_2|^2\rangle$ inside the shell is a quantity sensitive to interference between different electromagnetic modes, i.e., $f_{\ell}v_{\ell}^*$ and $g_{\ell}w_{\ell}^*$.
This is due to the interference between partial waves generated from Bessel or Neumann functions within the spherical shell.
Indeed, one can show that these interferences are related to the unconventional Fano resonance observed in the total scattering cross section~\cite{Arruda_PhysRevA87_2013}.  

In Fig.~\ref{fig4}, we show how the electric field intensity $\langle |\mathbf{E}|^2\rangle$ depends on the distance from the center of the sphere to its surface $r=b$.
We study three main frequencies obtained from $\sigma_{\rm sca}$ plotted in Fig.~\ref{fig2}: the dipole resonance ($\omega\approx 0.170\omega_{\rm p}$), the Fano dip ($\omega\approx 0.175\omega_{\rm p}$), and the quadrupole resonance ($\omega\approx 0.208\omega_{\rm p}$).
We verify that even at the Fano dip (with $\sigma_{\rm sca}\approx 0$) it is possible to obtain a large field intensity enhancement inside the (Si) core-shell (Ag) nanosphere.
Indeed, the intensity inside the lossless dielectric core ($r<a$) is even greater than the intensity obtained for the quadrupole resonance, which characterizes an off-resonance field enhancement at the subwavelength scale.

In the following, we use the ideas presented in this section to study how the Fano resonances are connected to the enhancement or suppression of the spontaneous-emission rate of optical emitters near plasmonic nanostructures.

\section{Spontaneous emission of a dipole emitter near a plasmonic nanoshell}
\label{sec3:Purcell}

Plasmonic surfaces are known to enhance or quench the fluorescence response of quantum emitters due to near- and far-field interactions between emitter and surface~\cite{Kerker_AppOpt19_1980,Wildea_SurfSciRep70_2015}.  
This modification of the spontaneous-emission rate of a quantum emitter due to the electromagnetic environment is generally refereed to as the Purcell effect~\cite{Kivshar_SciRep5_2015}.
Historically, this effect was first described by E.M. Purcell in the context of nuclear magnetic resonance~\cite{Purcell_PhysRev69_1946}, and was followed by the reports of K.H. Drexhage on the effects of metallic surfaces on fluorescence decay rate~\cite{Drexhage_JLum12_1970} and R.R. Chance et al. concerning molecular fluorescence near interfaces~\cite{Chance_JChemPhys63_1975}.
At present, this effect is widely used in several applications involving the enhancement and controlling of light emission and absorption at nanoscale, such as nanoplasmonic devices, nanoscale sensors, and the design of novel optical antennas in surface enhanced spectroscopy and microscopy~\cite{Kivshar_SciRep5_2015,Wildea_SurfSciRep70_2015,Novotny_PhysRevLett96_2006}.     

This section is devoted to the classical electrodynamics theory that describes the interaction between a single dipole emitter and a coated nanosphere.
In quantum electrodynamics, the standard approach to calculate the variation on linewidth and energy level shift of a quantum emitter due to boundary conditions is the first-order perturbation theory~\cite{Wylie_PhysRevA30_1984}.
In the weak coupling regime, the excited emitter decays exponentially to its ground state with life time $\tau=1/\Gamma$.
A remarkable feature of this approximation is that the decay rate $\Gamma$ of a quantum emitter in the vicinity of a body, normalized by the spontaneous-emission rate in free space $\Gamma_0$, can be calculated in the framework of classical electrodynamics~\cite{Wildea_SurfSciRep70_2015,Letokhov_JModOpt43_2_1996,Aguanno_PhysRevE69_2004,Zadkov_PhysRevA85_2012}.
In this case, the excited emitter is modeled as a point dipole source interacting with local electric field at the same position as the quantum emitter, and the Purcell factor $\Gamma/\Gamma_0$ is derived from the radiated power normalized to free space~\cite{Chew_JCPhys87_1987,Ruppin_JCPhys76_1982}.
This equivalence between classical and quantum calculations in the weak coupling regime occurs due to the fact that both the mode functions of the quantized electromagnetic fields and the classical electric field are derived from the same vector Helmholtz equation~\cite{Aguanno_PhysRevE69_2004,Milonni_Book1994}.

In the following, we present in Sec.~\ref{subsec3:Decay} an overview of the fully classical theory used to derive the spontaneous-emission rates of dipole emitters in close proximity of spheres.
In Sec.~\ref{subsec3:Efficiency}, we calculate the influence of near-field interactions on the radiation efficiency of a dipole emitter near a plasmonic nanoshell. 
The relation between Fano resonances and the spontaneous-emission rate is discussed in detail in Sec.~\ref{subsec3:Fano-Purcell}.
It is worth emphasizing that the final expressions for $\Gamma/\Gamma_0$ derived in Sec.~\ref{subsec3:Decay} agree with the first-order perturbation theory in the weak coupling regime~\cite{Chew_JCPhys87_1987,Kivshar_SciRep5_2015}.

\subsection{Radiative and non-radiative decay rates of a dipole emitter}
\label{subsec3:Decay}

\begin{figure}[htbp]
\centering
\includegraphics[width=.9\textwidth]{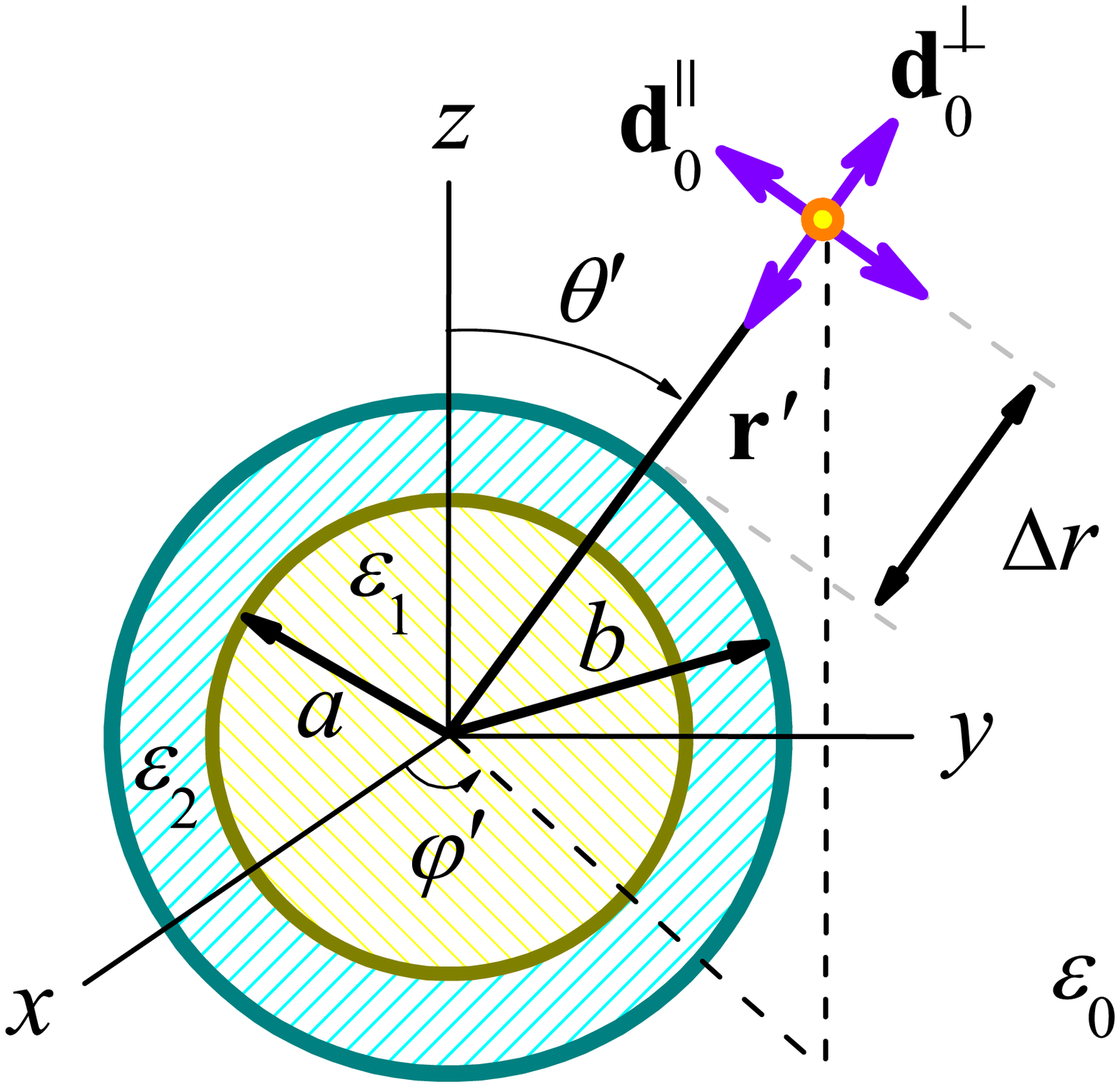}
\caption{An optical dipole emitter in the vicinity of a core-shell sphere in free space.
The inner sphere has radius $a$ and electric permittivity $\varepsilon_1$, whereas the outer sphere has radius $b$ and electric permittivity $\varepsilon_2$.
The surrounding medium is the vacuum $\varepsilon_0$.
The optical emitter is located at the position $\mathbf{r}'$, with $|\mathbf{r}'|=r'=b+\Delta r$. 
There are two basic orientations for the electric dipole moment $\mathbf{d}_0$ associated with the dipole emitter: it can be orthogonal ($\mathbf{d}_0^{\perp}$) or tangential ($\mathbf{d}_0^{||}$) to the spherical surface.
Any arbitrary dipole moment orientation in relation to the sphere can be decomposed in orthogonal and tangential contributions.}\label{fig5}
\end{figure}

Let us consider the same geometry investigated in Sec.~\ref{sec2:Mie}: a core-shell sphere of inner radius $a$ and outer radius $b$ in free space $(\varepsilon_0,\mu_0)$.
The sphere has optical properties $(\varepsilon_1,\mu_1)$ for the core $(r\leq a)$ and $(\varepsilon_2,\mu_2)$ for the shell $(a\leq r\leq b)$, as depicted in Fig.~\ref{fig5}.
Both core and shell consisting of isotropic and linear materials, and may have absorption and dispersion that satisfies the Kramers-Kronig relations~\cite{Farina_PhysRevA87_2013,Welsch_PhysRevA64_2001}.
In addition, we consider a single dipole emitter located at position $\mathbf{r}'$, with $r'=|\mathbf{r}'|>b$.
The dipole emitter is characterized by its electric dipole moment $\mathbf{d}_0$ and its emission frequency $\omega$.
The electric field emitted by this electric dipole in the region $b<r<r'$ can be expanded in terms of vector spherical harmonics~\cite{Chew_JCPhys87_1987,Ruppin_JCPhys76_1982} and reads

\begin{eqnarray}
\mathbf{E}_{\rm dip}^{\mathbf{d}_0}(r,\theta,\varphi)
&=&\sum_{\ell=1}^{\infty}\sum_{m=-\ell}^{\ell}\frac{1}{\ell(\ell+1)}\bigg\{\alpha_{\ell m}\frac{1}{k}{\nabla}\times\left[j_{\ell}(kr){\hat\mathbf{L}}Y_{\ell m}(\theta,\varphi)\right]\nonumber\\
&&+\beta_{\ell m}j_{\ell}(kr){\hat\mathbf{L}}Y_{\ell m}(\theta,\varphi)\bigg\},\label{E-dip}\\
\alpha_{\ell m}&=&-\imath k^2\mathbf{d}_{0}\cdot{\nabla}'\times\left[h_{\ell}^{(1)}(kr'){\hat\mathbf{L}}'Y_{\ell m}^*(\theta',\varphi')\right],\label{alpha}\\
\beta_{\ell m}&=&-\imath k^3 h_{\ell}^{(1)}(kr')\mathbf{d}_{0}\cdot\hat{\mathbf{L}}'Y_{\ell m}^*(\theta',\varphi'),\label{beta}
\end{eqnarray}
where $k=\omega\sqrt{\varepsilon_0\mu_0}$, $Y_{\ell m}(\theta,\varphi)$ is the spherical harmonics, and $\hat{\mathbf{L}}=-\imath{\mathbf{r}}\times{\nabla}$ is the angular momentum operator~\cite{Bohren_Book_1983}.
The derivation of Eq.~(\ref{beta}) can be found in Ref.~\cite{Letokhov_JModOpt43_1_1996}.
Here, the superindex $\mathbf{d}_0$ is just a reminder that the emitted electromagnetic fields depend on the dipole orientation.
Also, $\mathbf{E}_{\rm dip}^{\mathbf{d}_0}(\mathbf{r})$ for $r>r'$ can be readily obtained from Eqs.~(\ref{E-dip})--(\ref{beta}) by interchanging $j_{\ell}$ with $h_{\ell}^{(1)}$.
Here, the choice of a Hankel function of the first kind $h_{\ell}^{(1)}$ for outgoing waves is closely related to the assumption of a time harmonic dependence $e^{-\imath\omega t}$~\cite{Bohren_Book_1983}.
From Maxwell's curl equations, this implies a magnetic field $\mathbf{H}_{\rm dip}^{\mathbf{d}_0}=-\imath\nabla\times\mathbf{E}_{\rm dip}^{\mathbf{d}_0}/\omega\mu_0$.

The electromagnetic wave $[\mathbf{E}_{\rm dip}^{\mathbf{d}_0}(\mathbf{r}),\mathbf{H}_{\rm dip}^{\mathbf{d}_0}(\mathbf{r})]e^{-\imath\omega t}$ impinges on a spherical particle centered at $r=0$, with radius $b$, and it is scattered to the far field for $r>b$.
From the boundary conditions, one can show that the scattered electric field $\mathbf{E}_{\rm sca}^{\mathbf{d}_0}(\mathbf{r})$ can be obtained from Eq.~(\ref{E-dip}) by simply replacing coefficients $(\alpha_{\ell m},\beta_{\ell m})$ with  $(a_{\ell m},b_{\ell m})$ and the function $j_{\ell}$ with $h_{\ell}^{(1)}$~\cite{Chew_JCPhys87_1987,Letokhov_JModOpt43_2_1996,Ruppin_JCPhys76_1982,Letokhov_JModOpt43_1_1996}.
This procedure leads to
\begin{eqnarray}
\mathbf{E}_{\rm sca}^{\mathbf{d}_0}(r,\theta,\varphi)
&=&\sum_{\ell=1}^{\infty}\sum_{m=-\ell}^{\ell}\frac{1}{\ell(\ell+1)}
\bigg\{ a_{\ell m}\frac{1}{k}{\nabla}\times\left[h_{\ell}^{(1)}(kr){\hat\mathbf{L}}Y_{\ell m}(\theta,\varphi)\right]\nonumber\\
&&+b_{\ell m}h_{\ell}^{(1)}(kr){\hat\mathbf{L}}Y_{\ell m}(\theta,\varphi)
\bigg\},\label{E-sca}\\
a_{\ell m}&=&-\alpha_{\ell m}a_{\ell},\label{amn}\\
b_{\ell m}&=&-\beta_{\ell m}b_{\ell},\label{bmn}
\end{eqnarray}
where $\alpha_{\ell m}$ and $\beta_{\ell m}$ are given by Eqs.~(\ref{alpha}) and (\ref{beta}), respectively.
The coefficients $a_{\ell}$ and $b_{\ell}$, which encode the dependence on the sphere parameters, are the usual electric and magnetic Lorenz-Mie coefficients, respectively, given by Eqs.~(\ref{an}) and (\ref{bn}).

Using the Green's tensor formalism~\cite{Chew_JCPhys87_1987,Letokhov_JModOpt43_2_1996}, the solution for the total decay rate associated with an electric dipole moment ${\mathbf{d}}_0$ can be expressed as 
\begin{eqnarray}
\frac{\Gamma_{\rm total}^{\mathbf{d}_0}}{\Gamma_{0}}=1+\frac{6\pi\varepsilon_0}{k^3d_0^2}{\rm Im}\left[\mathbf{d}_0\cdot\mathbf{E}_{\rm sca}^{\mathbf{d}_0}(\mathbf{r}')\right],\label{equiv1}
\end{eqnarray}
where the scattered electric field contains the information of the environment in which the optical emitter is embedded.
Equation~(\ref{equiv1}) takes into account both radiative and non-radiative contributions, and provides a fully classical computational method to derive a quantum property of a system~\cite{Chew_JCPhys87_1987,Aguanno_PhysRevE69_2004}.
Physically, it is derived from the total power delivered by the optical emitter to the environment~\cite{Kivshar_SciRep5_2015}, $P_{\rm total}=-\omega{\rm Im}\{\mathbf{d}_0\cdot[\mathbf{E}_{\rm dip}^{\mathbf{d}_0}(\mathbf{r}')+\mathbf{E}_{\rm sca}^{\mathbf{d}_0}(\mathbf{r}')]\}/2$.

Now, let us now consider two basic orientations for the electric dipole moment in spherical geometry:
\begin{equation}
\mathbf{d}_0^{\perp} = d_0\hat{\bf e}_r\ ,\quad \mathbf{d}_0^{||} = \frac{d_0}{\sqrt{2}}\left(\hat{\bf e}_{\theta}+\hat{\bf e}_{\varphi}\right),
\end{equation}
where $\mathbf{d}_0^{||}$ is chosen for convenience~\cite{Arruda_PhysRevA96_2017}.
Without loss of generality, we set the position of the dipole emitter along the positive $z$-axis, i.e., $r'=z>b$ and $\theta'=\varphi'=0$.
As a result, since $P_{\ell}^{m}(\cos\theta')\propto \sin^m\theta'$ as $\theta'\to0$, only the terms with $m=0,\pm1$ contribute to the sum in Eqs.~(\ref{E-dip}) and (\ref{E-sca})~\cite{Letokhov_JModOpt43_2_1996}.
Substituting Eq.~(\ref{E-sca}) into Eq.~(\ref{equiv1}) for $\theta'=\varphi'=0$, we obtain the total decay rates associated with a dipole moment oriented orthogonal ($\mathbf{d}_0^{\perp})$ or tangential $(\mathbf{d}_0^{||})$ to the spherical surface, respectively:
\begin{eqnarray}
\frac{\Gamma_{\rm total}^{\perp}(kr')}{\Gamma_{0}}&=&1-\frac{3}{2}\sum_{\ell=1}^{\infty}\ell(\ell+1)(2\ell+1){\rm Re}\left\{a_{\ell}\left[\frac{h_{\ell}^{(1)}(kr')}{kr'}\right]^2\right\},\label{Gamma-perp}\\
\frac{\Gamma_{\rm total}^{||}(kr')}{\Gamma_{0}}&=&1-\frac{3}{4}\sum_{\ell=1}^{\infty}(2\ell+1){\rm Re}\left\{a_{\ell}\left[\frac{\xi_{\ell}'(kr')}{kr'}\right]^2+ b_{\ell} h_{\ell}^{(1)}(kr')^2\right\}.\label{Gamma-para}
\end{eqnarray}
For an electric dipole moment with arbitrary orientation in relation to the spherical surface, one can assume the spatial mean~\cite{Gaponenko_JPhysChem116_2012}: $\Gamma_{\rm total}= (\Gamma_{\rm total}^{\perp} + 2\Gamma_{\rm total}^{||})/3$.

Equations~(\ref{Gamma-perp}) and (\ref{Gamma-para}) contain both radiative and non-radiative contributions to the spontaneous-emission rate~\cite{Kivshar_SciRep5_2015}.
It is convenient to investigate these two contributions separately as they play different roles in near- and far-field interactions~\cite{Dujardin_OptExp16_2008}.
Indeed, for plasmonic spheres, the non-radiative contribution is related to an efficient coupling to surface plasmon modes in the near field.
Conversely, the radiative decay rate is associated with the excitation of Mie resonances in the far field.

In classical electrodynamics, the radiative decay rate $\Gamma_{\rm rad}^{\mathbf{d}_0}/\Gamma_0$ of a dipole emitter at the position $\mathbf{r}'$ is calculated via the total radiated power in the presence of the sphere normalized to free space~\cite{Ruppin_JCPhys76_1982}.
It can be calculated by integrating the radial component of the Poynting vector at the far field $(r\to\infty)$: $P_{\rm rad}=r^2\int {\rm d}\Omega\mathbf{S}\cdot\hat{\bf e}_r\propto r^2\int_{-1}^1 {\rm d}(\cos\theta)\int_{0}^{2\pi} {\rm d}\varphi|\mathbf{E}_{\rm dip}^{\mathbf{d}_0}(\mathbf{r})+\mathbf{E}_{\rm sca}^{\mathbf{d}_0}(\mathbf{r})|^2$, where $\mathbf{E}_{\rm dip}^{\mathbf{d}_0}(\mathbf{r})$ and $\mathbf{E}_{\rm sca}^{\mathbf{d}_0}(\mathbf{r})$ are defined in Eqs.~(\ref{E-dip}) and (\ref{E-sca}), respectively.
As a final result, we have
\begin{eqnarray}
\frac{\Gamma_{\rm rad}^{\perp}(kr')}{\Gamma_{0}}&=&\frac{3}{2}\sum_{\ell=1}^{\infty} \ell(\ell+1)(2\ell+1)\left|\frac{j_{\ell}(kr')-a_{\ell}h_{\ell}^{(1)}(kr')}{kr'}\right|^2,\label{Gamma-perprad}\\
\frac{\Gamma_{\rm rad}^{||}(kr')}{\Gamma_{0}}&=&\frac{3}{4}\sum_{\ell=1}^{\infty}(2\ell+1)\Bigg[\left|\frac{\psi_{\ell}'(kr')-a_{\ell}\xi_{\ell}'(kr')}{kr'}\right|^2\nonumber\\
&&+\left|j_{\ell}(kr')-b_{\ell}h_{\ell}^{(1)}(kr')\right|^2\Bigg].\label{Gamma-pararad}
\end{eqnarray}
For a detailed calculation of these expressions by using the Poynting vector, the interested reader is referred to Ref.~\cite{Chew_JCPhys87_1987}.
A different approach is discussed by Arruda et al.~\cite{Arruda_PhysRevA96_2017} using the Lorenz-Mie theory, in which the radiative decay rate is calculated straightforwardly from the intensity enhancement factor.
Indeed, one can verify that $\Gamma_{\rm rad}^{\mathbf{d}_0}(r')/\Gamma_0=\langle |\mathbf{d}_0\cdot[\mathbf{E}_{\rm in}(\mathbf{r}')+\mathbf{E}_{\rm sca}(\mathbf{r}')]|^2\rangle/\langle |\mathbf{d}_0\cdot\mathbf{E}_{\rm in}(\mathbf{r}')|^2\rangle$, where $\mathbf{E}_{\rm in}(\mathbf{r})$ and $\mathbf{E}_{\rm sca}(\mathbf{r})$ are given by Eqs.~(\ref{Ein}) and (\ref{Esca}), respectively, and $\langle \cdots \rangle = (1/{4\pi})\int_{0}^{4\pi}\Omega (\cdots)$ is the angle average~\cite{Arruda_PhysRevA96_2017}.
Once again, assuming the dipole has no defined orientation in space, one has from Eqs.~(\ref{Gamma-perprad}) and (\ref{Gamma-pararad}) the spatial mean $\Gamma_{\rm rad} = (\Gamma_{\rm rad}^{\perp} + 2\Gamma_{\rm rad}^{||})/3$.
In addition, by subtracting Eqs.~(\ref{Gamma-perprad}) and (\ref{Gamma-pararad}) from Eqs.~(\ref{Gamma-perp}) and (\ref{Gamma-para}), respectively, we finally obtain the non-radiative decay rates
\begin{eqnarray}
\frac{\Gamma_{\rm nrad}^{\perp}(kr')}{\Gamma_{0}}&=&\frac{3}{2}\sum_{\ell=1}^{\infty}
\ell(\ell+1)(2\ell+1)\left|\frac{h_{\ell}^{(1)}(kr')}{kr'}\right|^2{\rm Re}\left(a_{\ell}-|a_{\ell}|^2\right),\label{Gamma-perpn}\\
\frac{\Gamma_{\rm nrad}^{||}(kr')}{\Gamma_{0}}&=&\frac{3}{4}\sum_{\ell=1}^{\infty}(2\ell+1){\rm Re}\Bigg\{\left|\frac{\xi_{\ell}'(kr')}{kr'}\right|^2\left(a_{\ell}-|a_{\ell}|^2\right)\nonumber\\
&&+\left|h_{\ell}^{(1)}(kr')\right|^2\left(b_{\ell}-|b_{\ell}|^2\right)\Bigg\}.\label{Gamma-paran}
\end{eqnarray}

Although we have been discussing the case of an optical emitter with electric dipole radiation in the vicinity of a sphere, analogous expressions can be readily obtained for a magnetic dipole transition by interchanging $a_{\ell}$ with $b_{\ell}$~\cite{Chew_JCPhys87_1987}.

\subsection{Decay rates and radiation efficiency near a plasmonic nanoshell}
\label{subsec3:Efficiency}

The theory presented above is general and can be applied to arbitrary non-optically active spheres and single dipole emitters (quantum dots, atoms or molecules) in the weak coupling regime~\cite{Gaponenko_JPhysChem116_2012,Dujardin_OptExp16_2008,Bach_ACSNano7_2013}.
Here, we consider a realistic system for a dipole emitter near a plasmonic core-shell sphere composed of a silicon (Si) core and a silver (Ag) nanoshell.
We are interested in a configuration where the presence of a dielectric core strongly modifies the scattering response of a plasmonic nanoshell~\cite{Arruda_PhysRevA92_2015,Arruda_PhysRevA94_2016,Arruda_JOpSocAmA31_2014}, ultimately leading to unconventional Fano resonances~\cite{Arruda_PhysRevA96_2017,Halas_NanoLett10_2010}.
The optical and geometric parameters are the same of Sec.~\ref{sec2:Mie}: a dielectric (Si) core of refractive index $n_1=3.5$ and radius $a=60$~nm coated with a plasmonic (Ag) nanoshell of radius $b=90$~nm.

\begin{figure}[htbp]
\centering
\includegraphics[width=.95\textwidth]{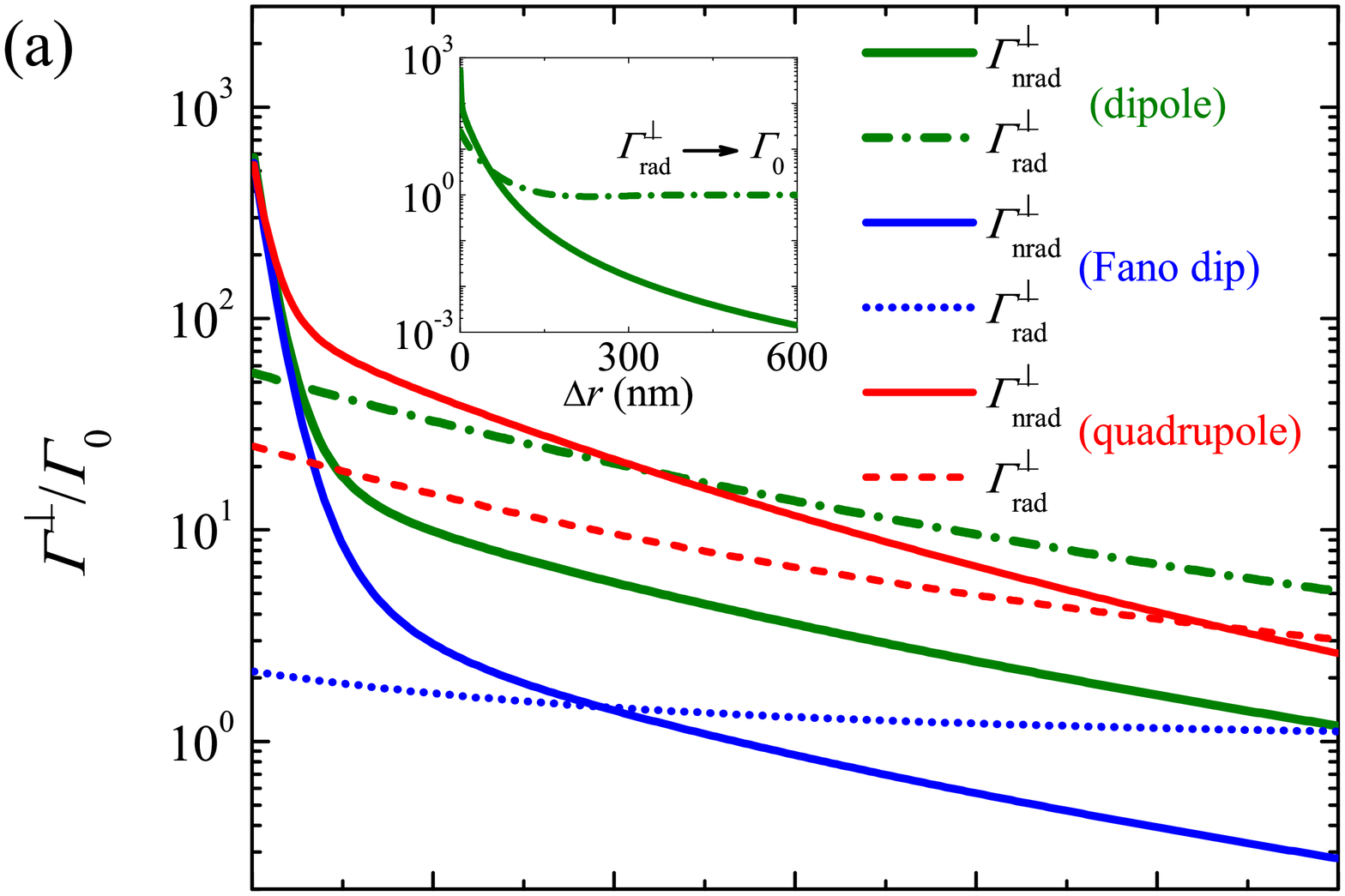}\vspace{-2.4cm}
\includegraphics[width=.95\textwidth]{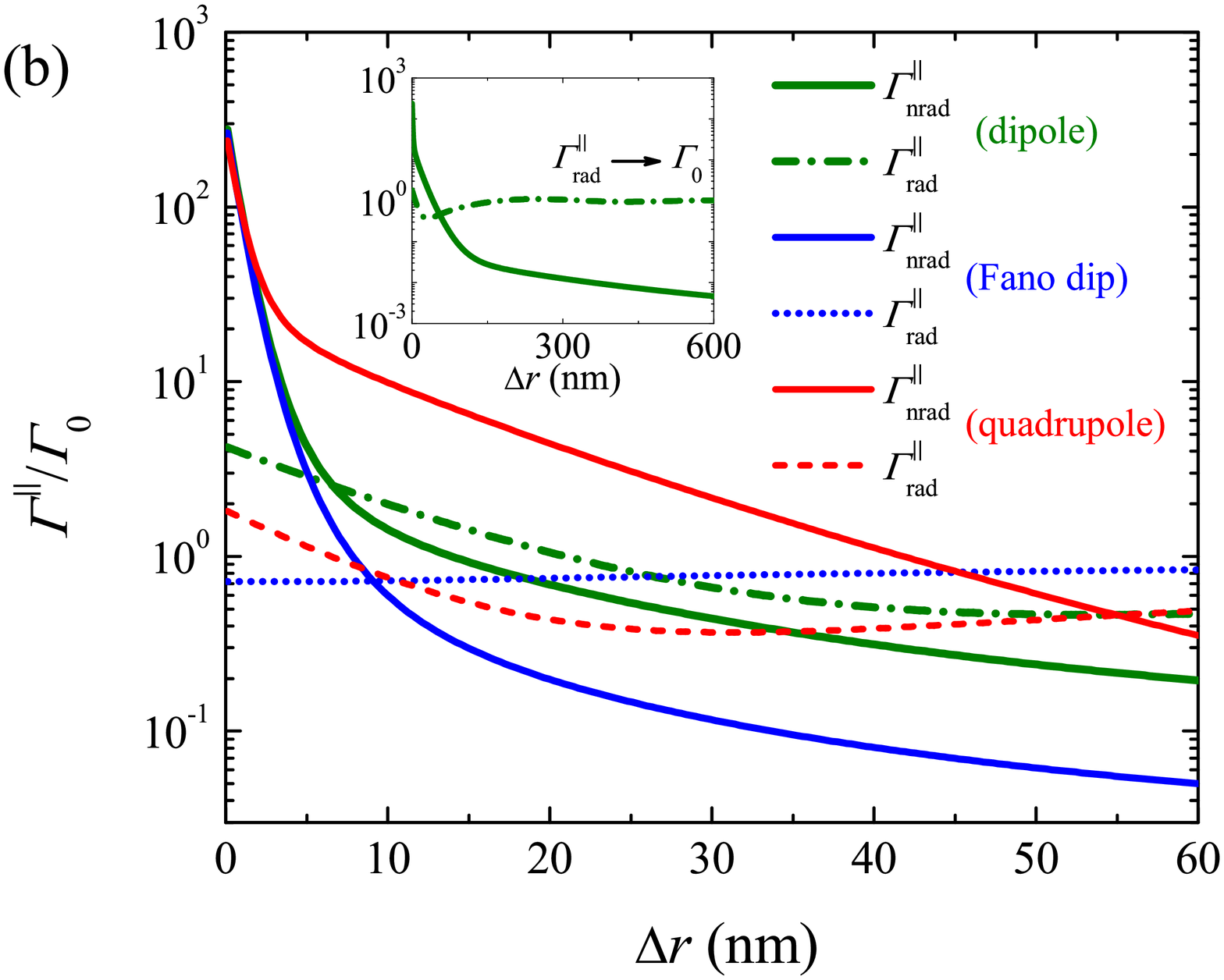}
\caption{Spontaneous decay rates $\Gamma$ of an optical dipole emitter near a (Si) core-shell (Ag) nanosphere in free space as a function of the distance $\Delta r$ between emitter and sphere.
The dielectric core has radius $a=60$~nm and refractive index $n_1=3.5$, and the Ag shell has radius $b=90$~nm and electric permittivity $\varepsilon_2=\varepsilon_{\rm Ag}(\omega)$ [Eq.~(\ref{eps-Ag})].
The decay rates are normalized by the corresponding decay rate $\Gamma_0$ in vacuum.
We consider three main frequencies obtained from Fig.~\ref{fig2}: dipole scattering resonance ($\omega\approx0.170\omega_{\rm p}$), Fano dip $(\omega\approx0.175\omega_{\rm p})$, and quadrupole resonance $(\omega\approx0.208\omega_{\rm p})$.  
The plots show radiative $(\Gamma_{\rm rad})$ and non-radiative $(\Gamma_{\rm nrad})$ decay rates associated with a point dipole oriented orthogonal (a) or parallel (b) to the spherical surface as a function of $\Delta r$.
The non-radiative decay rates dominate for $\Delta r\approx 0$ $(\Gamma_{\rm rad}\ll\Gamma_{\rm nrad})$.
The inset shows that $\Gamma_{\rm rad}\to\Gamma_0$ and $\Gamma_{\rm nrad}\to0$ for $\Delta r\gg b$ (far field).
At the Fano dip, $\Gamma_{\rm rad}\approx\Gamma_0$ irrespective of $\Delta r$ and dipole orientation.
}\label{fig6}
\end{figure}

Figure~\ref{fig6} shows the Purcell factor $\Gamma/\Gamma_0$ related to a single dipole emitter near a plasmonic shell as a function of the distance $\Delta r$ for two basic dipole moment orientations: orthogonal [Fig.~\ref{fig6}(a)] or parallel [Fig.~\ref{fig6}(b)] to the spherical surface. 
Based on the scattering cross section $\sigma_{\rm sca}$ plotted in Fig.~\ref{fig2}, we investigate three main frequencies for light emission: dipole scattering resonance $(\omega\approx 0.170\omega_{\rm p})$, Fano dip ($\omega\approx0.175\omega_{\rm p}$), and quadrupole scattering resonance $(\omega\approx 0.208\omega_{\rm p})$, where $\omega_{\rm p}$ is the Ag plasmon frequency.

As can be observed, in the vicinity of the plasmonic nanoshell $(\Delta r\to 0)$, non-radiative channels always dominate over far-field radiative processes, leading to $\Gamma_{\rm rad}^{\perp(||)}\ll \Gamma_{\rm nrad}^{\perp(||)}$. 
In the present system, this effect is mainly associated with ohmic losses on the plasmonic surface.
However, as the distance $\Delta r$ between emitter and nanoshell increases, the non-radiative decay rate decreases faster than the radiative one.
At the far field $(\Delta r\gg b)$, this results in $ \Gamma_{\rm rad}^{\perp(||)}\to \Gamma_0$ and $\Gamma_{\rm nrad}^{\perp(||)}\to 0$.

There are some interesting features in Fig.~\ref{fig6} that can be explained by light scattering theory.
For instance, the non-radiative decay rate $\Gamma_{\rm nrad}^{\perp(||)}$ associated with the quadrupole scattering resonance ($|a_2|^2$) is greater than that one related to the dipole scattering resonance ($|a_1|^2$).
This is an expected result, since the electric quadrupole scattering channel ($\ell=2$) is mainly associated with absorption, see Fig.~\ref{fig2}. 
In addition, note that the light emission at the Fano dip frequency leads to $\Gamma_{\rm rad}^{\perp(||)}\approx\Gamma_0$ irrespective of the distance $\Delta r$ between emitter and sphere.
Indeed, for a non-dissipative nanoshell, the net spontaneous-emission rate can be identically reduced to its vacuum value depending on the geometrical parameters of the plasmonic coating~\cite{Farina_PhysRevA87_2013}.
This effect is explained by the plasmonic cloaking of the dielectric sphere~\cite{Alu_PhysRevE72_2005}, since $\sigma_{\rm sca}\approx 0$ at the Fano dip ($\omega\approx 0.175\omega_{\rm p}$).
However, observe that the plasmonic cloaking is effective only from a certain finite distance $\Delta r$ of the nanoshell due to unavoidable non-radiative contributions of higher order dark modes $(\ell>1)$ at the near field. 

\begin{figure}[htbp!]
\centering
\includegraphics[width=\textwidth]{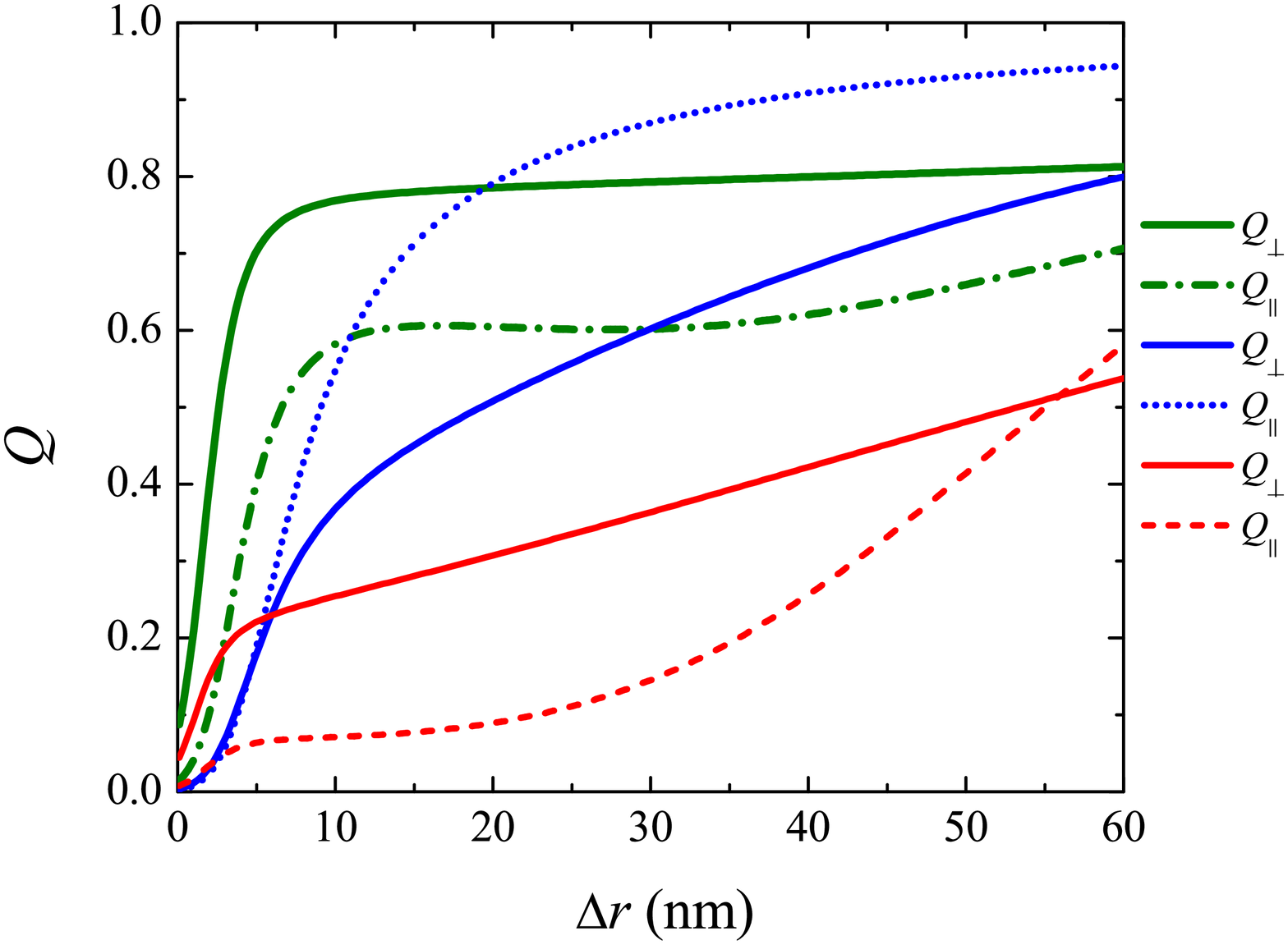}
\caption{Radiation efficiency $Q=\Gamma_{\rm rad}/\Gamma_{\rm total}$ associated with a dipole emitter in the vicinity of a plasmonic nanoshell in free space.
The system is composed of a (Si) core-shell (Ag) nanosphere with inner radius $a=60$~nm and outer radius $b=90$~nm.
The plots are calculated for an electric dipole moment $\mathbf{d}_0$ oriented orthogonal ($Q_{\perp}$) or tangential ($Q_{||}$) to the nanoshell as a function of distance $\Delta r$ for three frequencies obtained from Fig.~\ref{fig2}: dipole scattering resonance ($\omega\approx0.170\omega_{\rm p}$, solid and dash-dotted green lines), Fano dip ($\omega\approx0.175\omega_{\rm p}$, solid and dotted blue lines), and quadrupole resonance ($\omega\approx0.208\omega_{\rm p}$, solid and dashed red lines).
For $\Delta r\approx 0$ or $\Delta r\gg b$, one has $Q_{\perp(||)}\to 0$ or $Q_{\perp(||)}\to1$, respectively.}\label{fig7}
\end{figure}

To clarify the role of radiative and non-radiative contributions on the spontaneous-emission rate of an optical emitter, it is convenient to define the radiation efficiency of the light emission.
The radiation efficiency $Q$ of an emitter with negligible internal losses is defined as~\cite{Kivshar_SciRep5_2015}
\begin{equation}
Q_{\mathbf{d}_0}(kr')=\frac{\Gamma_{\rm rad}^{\mathbf{d}_0}(kr')}{\Gamma_{\rm rad}^{\mathbf{d}_0}(kr')+\Gamma_{\rm nrad}^{\mathbf{d}_0}(kr')},\label{Q-eff}
\end{equation}
where the corresponding radiative and non-radiative decay rates are calculated in Sec.~\ref{subsec3:Decay}.
Using Eq.~(\ref{Q-eff}), we plot in Fig.~\ref{fig7} the competition between far-field radiation and ohmic losses on the surface of the plasmonic nanoshell as a function of $\Delta r$.
As expected, the radiation efficiency $Q$ for both dipole moment orientations vanishes at the plasmonic surface $(\Delta r\approx 0$).
In particular, note that $Q_{\perp}>Q_{||}$ in general, which means a more efficient coupling between the electric dipole moment $\mathbf{d}_0$ oriented orthogonal to the spherical surface than the parallel orientation. 
Among the chosen light emission frequencies, the lowest values of efficiency at the near field is obtained for the quadrupole scattering resonance frequency $\omega\approx0.208\omega_{\rm p}$. 

\subsection{The Purcell effect and Fano resonances in plasmonic nanoshells}
\label{subsec3:Fano-Purcell}

The influence of Fano resonances on the Purcell factor is revealed when one considers $\Gamma/\Gamma_0$ as a function of the light emission frequency $\omega$~\cite{Kivshar_SciRep5_2015}.
Recently, it has been analytically demonstrated that the fluorescence enhancement of dipole emitter near a plasmonic nanoshell as a function of the excitation frequency also exhibits an asymmetric Fano line shape~\cite{Arruda_PhysRevA96_2017}.
Here, we use the same arguments applied in Ref.~\cite{Arruda_PhysRevA96_2017} to describe the Fano effect on the Purcell factor of a dipole emitter in close proximity of plasmonic nanoshells.
We focus only on the radiative contribution since we are interested in the dipole mode $(\ell=1)$ excited in the sphere, which is related to the unconventional Fano resonance.
For the non-radiative contribution $\Gamma_{\rm nrad}^{\perp(||)}$, the quadrupole mode ($\ell=2$) excited in the particle dominates the spectrum with a Lorentzian line shape, whereas higher order dark modes ($\ell>2$) contribute to $\Gamma_{\rm nrad}^{\perp(||)}$ in the near field, leading to a broad spectral line ($Q\to0$, see Fig.~\ref{fig7}).
This influence of higher dark modes is the main reason why the dipole approximation fails to describe near-field interactions between an optical emitter and a plasmonic nanosphere~\cite{Novotny_PhysRevLett96_2006}.
Conversely, since $\Gamma_{\rm rad}^{\perp(||)}$ is related to the far-field radiation, we can restrict our discussion to $\ell=1$ for $kb< 1$ and $k\Delta r< 1$ in the vicinity of the dipole scattering resonance $(|a_1|^2)$.
All the numerical calculations, however, are performed with the exact expressions derived in Sec.~\ref{subsec3:Decay}. 

\begin{figure}[htbp]
\centering
\includegraphics[width=.9\textwidth]{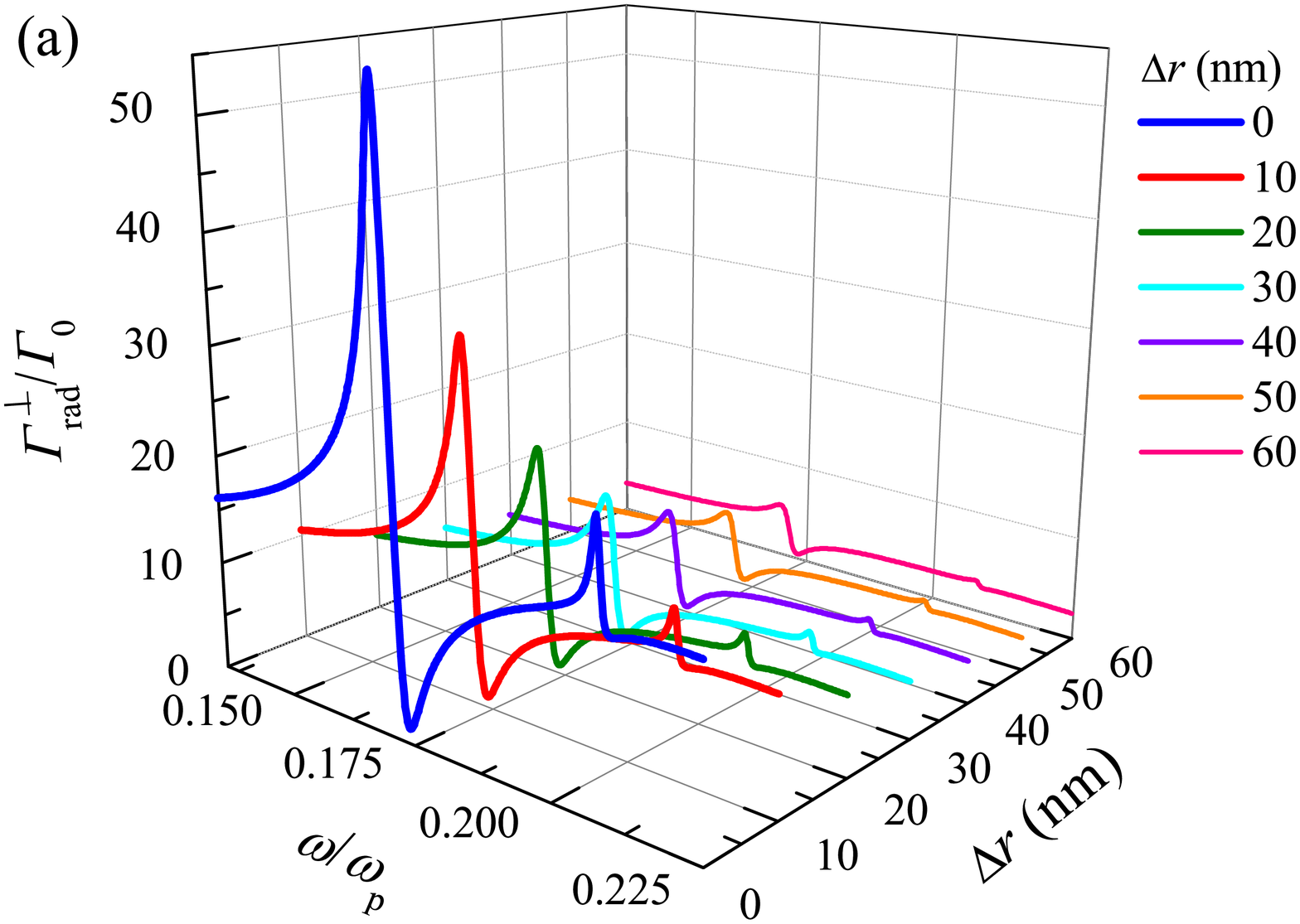}\vspace{-.5cm}
\includegraphics[width=.9\textwidth]{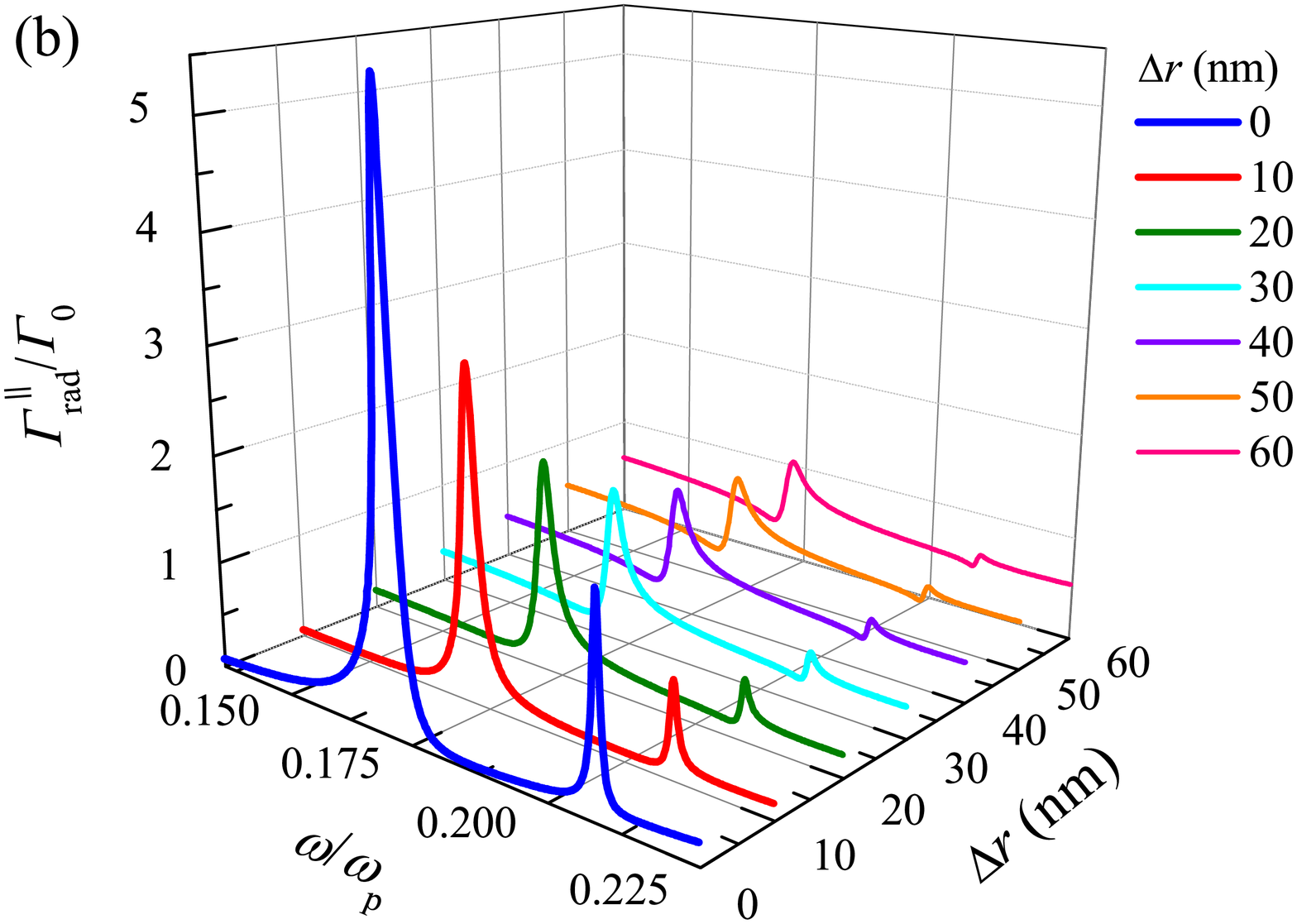}
\caption{Radiative decay rates $\Gamma_{\rm rad}$ of a dipole emitter near a (Si) core-shell (Ag) nanoparticle as a function of the light emission frequency $\omega$.
We consider several distances $\Delta r$ between emitter and coated sphere, which has inner radius $a=60$~nm and outer radius $b=90$~nm.
(a) The plot shows the radiative decay rate $\Gamma_{\rm rad}^{\perp}$ of a dipole emitter with orthogonal orientation in relation to the spherical shell.
For $\omega\approx0.170\omega_{\rm p}$ (dipole scattering resonance) and $\omega\approx0.208\omega_{\rm p}$ (quadrupole scattering resonance), one has asymmetric Fano line shapes irrespective of the distance $\Delta r$.
(b) The plot shows $\Gamma_{\rm rad}^{||}$ of a dipole emitter with tangential orientation in relation to the spherical surface.
For $\Delta r\approx 0$, one has symmetric Lorentzian line shapes.
From $\Delta r >10$~nm, these Lorentzian line shapes change to Fano line shapes.
}\label{fig8}
\end{figure}

In Fig.~\ref{fig8}, we plot $\Gamma_{\rm rad}^{\perp}$ and $\Gamma_{\rm rad}^{||}$ as a function of the light emission frequency $\omega$, and for several distances $\Delta r$ between emitter and plasmonic surface.
By comparing Fig.~\ref{fig8}(a) and Fig.~\ref{fig8}(b), we see clearly that $\Gamma_{\rm rad}^{\perp}$ is one order of magnitude greater than $\Gamma_{\rm rad}^{||}$, confirming that the coupling between emitter and plasmonic nanoshell is stronger for the orthogonal orientation of the dipole moment.
More importantly, on one hand, the plots of $\Gamma_{\rm rad}^{\perp}$ exhibit Fano resonances for both dipole ($\omega\approx 0.170\omega_{\rm p}$) and quadrupole $(\omega\approx 0.208\omega_{\rm p})$ modes irrespective of $\Delta r$.
On the other hand, the plots of $\Gamma_{\rm rad}^{||}$ exhibit symmetric Lorentzian profiles for $\Delta r\ll b$ and, as $\Delta r$ increases, it develops to a Fano line shape.

As discussed by Arruda et al.~\cite{Arruda_PhysRevA96_2017}, the Lorentzian line shape observed in $\Gamma_{\rm rad}^{||}(\omega)$ in the near field, that changes into a Fano line shape in the far field, is a consequence of the core-shell geometry. 
Physically, the electric dipole moment $\mathbf{d}_{0}^{||}$ associated with the optical emitter induces an oppositely directed dipole moment on the plasmonic nanoshell surface, with almost the same amplitude~\cite{Zadkov_PhysRevA85_2012}.
This interaction cancels out the broad dipole mode excited in the plasmonic sphere, but does not cancel out the narrow dipole mode $(\ell=1)$ at the plasmonic inner shell surface.
According to Refs.~\cite{Tribelsky_PhysRevA93_2016,Arruda_PhysRevA96_2017}, we can rewrite the electric Lorenz-Mie coefficient $a_{\ell}$, Eq.~(\ref{an}), as
\begin{eqnarray}
a_{\ell} = a_{\ell}^{\rm PEC} - \frac{\left[\psi_{\ell}'(n_2kb)g_{\ell} - \chi_{\ell}'(n_2kb)w_{\ell}\right]}{n_2\xi_{\ell}'(kb)},\label{a-ell} 
\end{eqnarray}
where $g_{\ell}$ and $w_{\ell}$ are the Lorenz-Mie coefficients of electromagnetic fields within the plasmonic shell, Eqs.~(\ref{gn}) and (\ref{wn}), respectively.

The first term in Eq.~(\ref{a-ell}) is the coefficient of a perfectly electric conducting (PEC) sphere $(n_2\to\infty)$: $a_{\ell}\to a_{\ell}^{\rm PEC}\equiv\psi_{\ell}'(kb)/\xi_{\ell}'(kb)$~\cite{Bohren_Book_1983}.
Here, this coefficient is related to the broad electric dipole mode $(\ell=1)$, while the second term accounts for the narrow electric dipole mode related to the plasmonic inner shell surface.
By inspection of Eqs.~(\ref{Gamma-perprad}) and (\ref{Gamma-pararad}), it is easily confirmed that the term $a_1^{\rm PEC}$ in Eq.~(\ref{a-ell}) is canceled out for $r'=b$ and $\ell=1$ only in $\Gamma_{\rm rad}^{||}$, leading to a Lorentzian line shape response as a function of frequency.
As the distance between the dipole and the nanoshell becomes greater, the influence of the broad dipole mode in the Purcell factor increases, leading to a Fano resonance.

\begin{figure}[htbp]
\centering
\includegraphics[width=\textwidth]{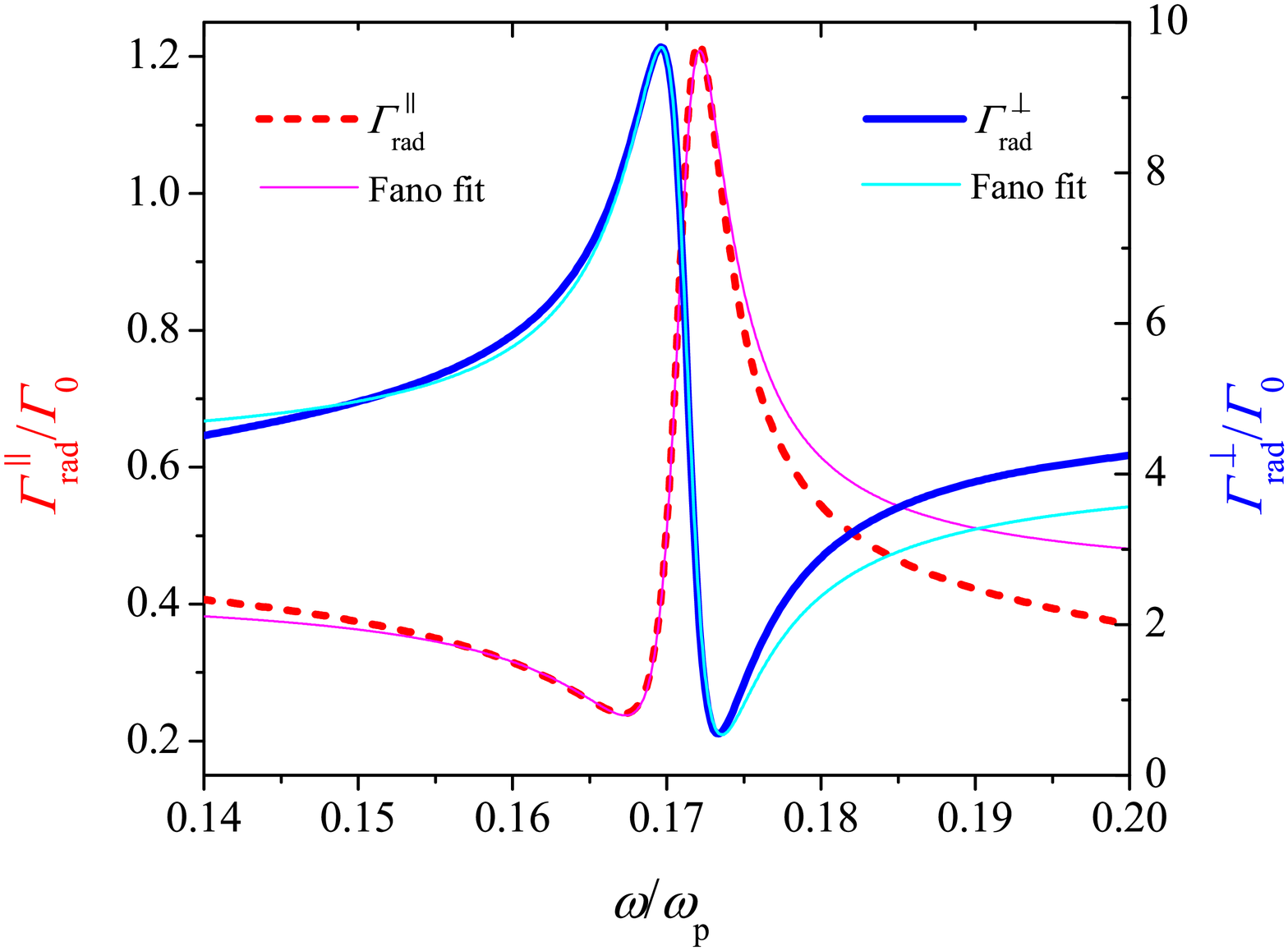}
\caption{Radiative decay rates related to an optical emitter located at $\Delta r=40$~nm from a (Si) core-shell (Ag) nanoparticle with inner radius $a=60$~nm and outer radius $b=90$~nm.
Both orthogonal $(\Gamma_{\rm rad}^{\perp})$ and parallel $(\Gamma_{\rm rad}^{||})$ orientations of the electric dipole moment $\mathbf{d}_0$ in relation to the spherical surface present a Fano ressonance around $\omega\approx0.170\omega_{\rm p}$, where $\omega_{\rm p}$ is the Ag plasmon frequency.
The corresponding Fano asymmetry parameters of the Purcell factors are $q_{\rm P}^{\perp}\approx-1.2$ and $q_{\rm P}^{||}\approx 2.0$.}\label{fig9}
\end{figure}

\begin{figure}[htbp]
\centering
\includegraphics[width=\textwidth]{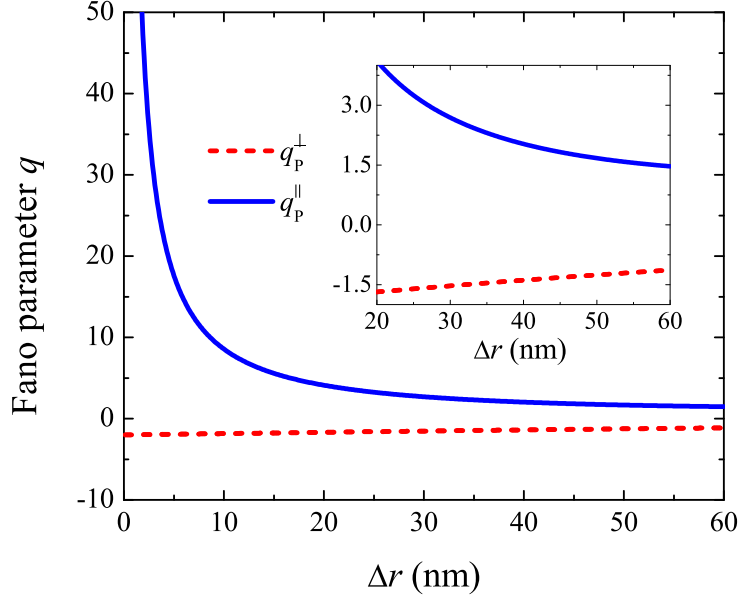}
\caption{Fano asymmetry parameters associated with the Purcell factor of an optical emitter in the vicinity of a (Si) core-shell (Ag) nanoparticle with light emission frequency $\omega=0.170\omega_{\rm p}$ (dipole scattering resonance).
The coated sphere has inner radius $a=60$~nm and outer radius $b=90$~nm.
The Fano parameters $q_{\rm P}^{\perp}$ and $q_{\rm P}^{||}$ are calculated from Eqs.~(\ref{q-perp}) and (\ref{q-para}) as a function of the distance $\Delta r$ between emitter and spherical surface.
For the electric dipole moment $\mathbf{d}_0$ oriented tangential to the spherical surface, we have $q_{\rm P}^{||}\to\infty$ as $\Delta r\to 0$.
The inset shows that $q_{\rm P}^{||}$ is finite for $\Delta r>20$~nm and has opposite sign in relation to $q_{\rm P}^{\perp}$.
These curves can be used to fit the plots in Fig.~\ref{fig8}.}\label{fig10}
\end{figure}

In Fig.~\ref{fig9}, we compare $\Gamma_{\rm rad}^{\perp}$ and $\Gamma_{\rm rad}^{||}$ for $\Delta r=40$~nm.
Both profiles present Fano line shapes, with Fano asymmetry parameters $q_{\rm P}^{\perp}$ and $q_{\rm P}^{\perp}$ with opposite sign.
These Fano parameters are related to the unconventional Fano resonance in the scattering cross section $\sigma_{\rm sca}$, where $q_{\rm LM}= {\chi_{1}'(kb)}/{\psi_{1}'(kb)}$ for $\ell =1$.
In particular, observe in Fig.~\ref{fig9} that the fitted Fano curves are better for low frequencies (large wavelengths).
Assuming the dipole approximation, i.e., $kr\ll1$ and $kb\ll1$, we obtain
\begin{eqnarray}
\frac{\Gamma_{\rm rad}^{\perp(||)}(\omega)}{\Gamma_0}\approx F_1^{\perp(||)}\left\{\frac{\left[\displaystyle\frac{\zeta'(\omega)}{\zeta''(\omega)+1} + q_{\rm P}^{\perp(||)}\right]^2 + \left[\displaystyle\frac{\zeta''(\omega)}{\zeta''(\omega)+1}\right]^2}{\left[\displaystyle\frac{\zeta'(\omega)}{\zeta''(\omega)+1}\right]^2 + 1}\right\},\label{G-approx}
\end{eqnarray}
where the prefactors for the two electric dipole orientations are
\begin{eqnarray}
F_1^{\perp} &=&\frac{9\left[j_1(kr')q_{\rm LM} + y_1(kr')\right]^2}{(kr')^2(1+q_{\rm LM}^2)},\\
F_1^{||} &=&\frac{9\left[\psi_1'(kr')q_{\rm LM} - \chi_1'(kr')\right]^2}{4(kr')^2(1+q_{\rm LM}^2)};
\end{eqnarray}
the corresponding Fano asymmetry parameters are 
\begin{eqnarray}
q_{\rm P}^{\perp} &=&\frac{1}{1+\zeta''(\omega)}\left[\frac{y_1(kr')q_{\rm LM}-j_1(kr')}{j_1(kr')q_{\rm LM}+y_1(kr')}\right],\label{q-perp}\\
q_{\rm P}^{||} &=&-\frac{1}{1+\zeta''(\omega)}\left[\frac{ \chi_1'(kr')q_{\rm LM}+\psi_1'(kr')}{\psi_1'(kr')q_{\rm LM}-\chi_1'(kr')}\right],\label{q-para}
\end{eqnarray}
where $\zeta(\omega)$ is defined in Eq.~(\ref{zeta}) and $r'=b+\Delta r$.

From Eqs.~(\ref{G-approx})--(\ref{q-para}), it becomes clear that only $|q_{\rm P}^{||}|\to\infty$ when $r'\to b$, which leads to a Lorentzian line shape in the near field for $\Gamma_{\rm rad}^{||}$.
This fact is shown explicitly in Fig.~\ref{fig10}, where we plot the corresponding Fano parameters that fit the plots in Fig.~\ref{fig8} by using Eqs.~(\ref{q-perp}) and (\ref{q-para}).
In particular, it is worth mentioning that Eqs.~(\ref{q-perp}) and (\ref{q-para}) can be easily generalized to an arbitrary $\ell$, since they are not approximate expressions.

\section{Conclusion}
\label{sec4:Conclusion}

Based on the complete Lorenz-Mie theory, we have investigated the role of Fano resonances in plasmonic core-shell spheres and their influence on the spontaneous-emission rate of optical emitters in close proximity of a nanoshell.
We have briefly discussed the appearance of conventional and unconventional Fano resonances in the light scattering by single-layered spheres.
Both resonances arise from the interference between electromagnetic modes excited in the particle and can be associated with the off-resonance field enhancement and saddle points in the energy flow around the particle.
For an optical emitter with dipole moment oriented tangentially to a plasmonic nanoshell, we have obtained a symmetric Lorentzian line shape response in the near field that changes into a Fano resonance in the far field, with Fano asymmetry parameter of opposite sign compared to the dipole moment oriented normally to the spherical surface.
This effect has been explained by the different role played by the induced electric dipole moment in the plasmonic nanoshell for both dipole moment orientations.
More importantly, we have unveiled the relation between Fano resonances in light scattering and the Purcell effect.
These analytical results shed light on a fundamental problem of Fano-like resonances in nanoplasmonics, and they may have interesting applications for enhancing and controlling the light emission and absorption of optical dipole emitters near metal-based nanostructures.

\begin{acknowledgement}
The authors thank John Weiner for the fruitful collaboration, discussions and suggestions to improve this study. 
T.J.A., R.B., and Ph.W.C. acknowledge S\~ao Paulo Research Foundation (FAPESP) (Grant Nos. 2015/21194-3, 2014/01491-0, and 2013/04162-5, respectively) for financial support.
A.S.M. holds grants from Conselho Nacional de Desenvolvimento Cient\'{\i}fico e Tecnol\'ogico (CNPq) (Grant No. 307948/2014-5).
F.A.P. acknowledges The Royal Society-Newton Advanced Fellowship (Grant No. NA150208), Coordena\c{c}\~ao de Aperfei\c{c}oamento de Pessoal de N\'ivel Superior (CAPES) (Grant No. BEX 1497/14-6),  Funda\c{c}\~ao Carlos Chagas Filho de Amparo \`a Pesquisa do Estado do Rio de Janeiro (FAPERJ) (Grant No. APQ1-210.611/2016), and CNPq (Grant No. 303286/2013-0) for financial support.
S.S. is supported by the Fulbright-Cottrell Award.
\end{acknowledgement}
%
%\section*{Appendix}
%\addcontentsline{toc}{section}{Appendix}
%
%
%When placed at the end of a chapter or contribution (as opposed to at the end of the book), the numbering of tables, figures, and equations in the appendix section continues on from that in the main text. Hence please \textit{do not} use the \verb|appendix| command when writing an appendix at the end of your chapter or contribution. If there is only one the appendix is designated ``Appendix'', or ``Appendix 1'', or ``Appendix 2'', etc. if there is more than one.

%%%%%%%%%%%%%%%%%%%%%%%% referenc.tex %%%%%%%%%%%%%%%%%%%%%%%%%%%%%%
% sample references
% %
% Use this file as a template for your own input.
%
%%%%%%%%%%%%%%%%%%%%%%%% Springer-Verlag %%%%%%%%%%%%%%%%%%%%%%%%%%
%
% BibTeX users please use
% \bibliographystyle{}
% \bibliography{}
%

\end{document}